\documentclass[lettersize,journal]{IEEEtran}
\usepackage{xcolor}
\usepackage{amsmath,amsfonts}
\usepackage{algorithmic}
\usepackage{array}
\usepackage[caption=false,font=normalsize,labelfont=sf,textfont=sf]{subfig}
\usepackage{textcomp}
\usepackage{hyperref}
\usepackage{stfloats}
\usepackage{url}
\usepackage{verbatim}
\usepackage{graphicx}
\usepackage{algorithm}
\usepackage{siunitx}
\usepackage{upgreek}
\usepackage{orcidlink}

\def\BibTeX{{\rm B\kern-.05em{\sc i\kern-.025em b}\kern-.08em
    T\kern-.1667em\lower.7ex\hbox{E}\kern-.125emX}}
\usepackage{balance}

\begin{document}
\title{When the UE Turns Adversary: Real-Time Uplink Jamming from Within 5G Networks}
\author{Rosolino~Alaimo\orcidlink{0009-0008-1319-0145},~
Alessandra~Dino\orcidlink{0009-0005-4869-8158},~
Ilenia~Tinnirello\orcidlink{0000-0002-1305-0248},~
and~Domenico~Garlisi\orcidlink{0000-0001-6256-2752}

\thanks{This work has been submitted to the IEEE for possible publication. Copyright may be transferred without notice, after which this version may no longer be accessible.}
}

\markboth{Journal of \LaTeX\ Class Files,~Vol.~18, No.~9, September~2020}%
{How to Use the IEEEtran \LaTeX \ Templates}

\maketitle

\begin{abstract}

This paper presents an investigation of a novel class of stealthy and selective reactive jamming attacks targeting the Physical Uplink Shared Channel (PUSCH) in 5G New Radio (NR) networks. We design and implement STORM-RJ (Stealthy Timing Obstruction and Radio Manipulation – Reactive Jamming), a Software-Defined Radio (SDR)-based adversarial framework that enables highly precise, time–frequency aligned interference by dynamically adapting the bandwidth and center frequency of injected noise bursts in real time.
STORM-RJ leverages decoded Downlink Control Information (DCI) to identify Uplink-Grants (UL-Grants) and synchronizes interference exactly with the resource blocks allocated to a target User Equipment (UE). 
We further characterize and mitigate the dominant latency sources — both at the software processing and hardware Radio Frequency (RF) frontend levels — to achieve a rapid jamming response upon grant detection.
We conduct a comparative analysis of high-level versus low-level radio control strategies, demonstrating that only low-level tuning provides the microsecond-scale responsiveness necessary to meet 5G-NR timing constraints for effective reactive jamming. 
We analyze the practical feasibility of such selective jamming under realistic hardware and timing constraints, highlighting key trade-offs between SDR flexibility, processing latency, and synchronization accuracy. Finally, we discuss potential mitigation strategies, including Hybrid Automatic Repeat reQuest (HARQ) anomaly detection.

\end{abstract}


\section{Introduction}
Modern telecommunication systems are characterized by an ever-increasing level of architectural complexity, driven by the demand for higher throughput, lower latency, and massive device connectivity. While these advancements enable a wide range of new applications, they also expand the potential attack surface of wireless infrastructures. Despite continuous improvements in protocol design and security mechanisms, undiscovered vulnerabilities may still remain within communication systems, posing risks to both network availability and the confidentiality of transmitted data.  Consequently, modern communication protocols must integrate advanced security mechanisms aimed at reducing exposure to malicious external attacks, a design principle that has guided the evolution of telecommunication systems from early architectures to contemporary 5G networks \cite{6815891}.
Among the various threats affecting wireless networks, jamming attacks represent one of the most critical concerns at the physical layer. By intentionally injecting interference into the wireless medium, an adversary can degrade link quality, reduce throughput, or completely disrupt communication services. While traditional jamming techniques rely on wideband interference that affects large portions of the spectrum, recent research has demonstrated the feasibility of more selective and energy-efficient attacks that exploit knowledge of the communication protocol to target specific transmissions. These techniques allow attackers to concentrate interference on carefully chosen time–frequency resources, significantly increasing the effectiveness of the attack while reducing its detectability \cite{harvanek2024survey}.
The 5G New Radio (5G-NR) architecture introduces new opportunities for such targeted attacks due to its highly dynamic and flexible resource allocation mechanisms. In particular, uplink transmissions are scheduled dynamically by the Next-Generation NodeB (gNB) through the Physical Downlink Control Channel (PDCCH), which carries the encoded Downlink Control Information (DCI) containing all scheduling parameters assigned to a specific UE. These messages specify both the time and frequency resources assigned to a User Equipment (UE). If an adversary gains access to this scheduling information, it becomes possible to align interference precisely with the scheduled uplink transmission, enabling highly selective reactive jamming strategies.
This study investigates a novel attack model in which the adversary leverages a compromised UE to extract decoded DCI and forward it to an external jamming device. Building upon this concept, we design and implement STORM-RJ (Stealthy Timing Obstruction and Radio Manipulation – Reactive Jamming), a Software-Defined Radio (SDR) framework capable of generating interference bursts that are precisely aligned with the bandwidth and central frequency of the targeted uplink transmission.
The present study extends the original STORM architecture \cite{STORM} by enabling highly selective interference of uplink transmissions through dynamic bandwidth adaptation and ultra-fast DSP-based frequency retuning.
This enables interference to be precisely aligned in time and frequency while matching the bandwidth of the dynamically scheduled uplink resources of the target UE.
A key challenge in implementing such reactive attacks lies in meeting the strict timing constraints imposed by the 5G-NR scheduling framework. In particular, the jammer must react within the short interval between the reception of the Uplink-Grant (UL-Grant) and the actual transmission of the UE. To address this challenge, this study investigates multiple engineering aspects that determine the practical feasibility of real-time reactive jamming, including latency sources in SDR platforms, frequency tuning strategies, and bandwidth-adaptive interference generation.
Specifically, this work makes the following contributions:
\begin{itemize}
    \item Design and implementation of STORM-RJ, an SDR-based framework capable of performing selective reactive jamming of 5G-NR uplink transmissions.
    
    \item Analysis of latency constraints affecting reactive jamming, including the delays associated with scheduling-information acquisition, Radio Frequency (RF) retuning and signal generation.
    
    \item Comparison of high-level and low-level frequency tuning strategies with different hardware, demonstrating that digital baseband tuning enables microsecond-scale frequency adjustments suitable for real-time operation.

    \item Analysis of throughput and Hybrid Automatic Repeat reQuest (HARQ) retransmission under attack conditions, demonstrating full channel degradation control.
    
    \item Discussion of potential countermeasures, including anomaly detection based on HARQ feedback patterns and physical-layer signal fingerprinting.
\end{itemize}
Through experimental validation using multiple SDR platforms, we demonstrate that selective uplink jamming aligned with dynamically scheduled resources is practically achievable under realistic hardware constraints.
The remainder of this paper is organized as follows. Section \ref{sec:related} reviews papers on the relative jamming attacks. Section \ref{sec:technical background} provides the technical background on the 5G-NR uplink resource structure and scheduling mechanisms relevant to the attack model. Section \ref{sec:methods} describes the experimental setup and the architecture of the proposed STORM-RJ framework, including the techniques used for dynamic bandwidth selection and frequency tuning. Section \ref{sec:results} presents the experimental results and evaluates the effectiveness of the proposed jamming approach under different hardware configurations. Section \ref{Defensive Considerations} discusses possible defensive strategies and mitigation techniques against this class of attacks. Finally, Section \ref{Conclusion and future work} concludes the paper and outlines directions for future research.

\section{Related Works}
\label{sec:related}
As discussed in \cite{9733393}, jamming represents one of the most critical and actively researched threats in Physical-Layer security for wireless communication systems. 
In terms of the 5G system, the vulnerabilities of 5G-NR channels and signals are qualitatively analyzed in the context of smart jamming attacks in \cite{9031175}. The study highlights how certain structural aspects of the 5G-NR standard may be exploited by adversaries. However, it does not include practical implementations or experimental validations of the proposed attack scenarios. The paper in \cite{10186886} developed a jammer capable of extracting the C-RNTIs of UEs connected to the network using a custom technique based on polar decoding lattices. This enables the attacker to decode UL-Grants and identify the specific uplink resources assigned to each UE.
Unlike that work, the approach presented in this paper relies on a different attack strategy, leveraging a compromised UE to obtain the UL-Grant information.
The work presented in \cite{STORM} introduces a system capable of disrupting the 5G Synchronization Signal Block (SSB), which UEs rely on for initial synchronization and cell selection during network access \cite{chen2020design}. It demonstrates a jamming approach that improves energy efficiency by confining the attack to a specific frequency bandwidth aligned with SSB resources.
Time and frequency domain synchronization are addressed in \cite{IEEE-COINS-2025}, which presents a preliminary version of the STORM framework. This version is capable of intercepting and selectively targeting payload data transmitted by the UE. The proposed approach includes traffic analysis mechanisms designed to extract the exact Transmission Time Interval ($\text{tti}_\text{tx}$) from the UL-Grant. This information allows the jammer to precisely inject white noise over the scheduled uplink resources, both in time and frequency. To achieve this level of precision, accurate synchronization is required; accordingly, the study focuses on identifying appropriate reference signals to align the attacker with the gNB timing. A quantitative evaluation of the resulting uplink throughput degradation is also provided.
It is important to emphasize that the jammer’s reaction time is a critical factor for the success of the attack. If either recognized mechanism fails to respond with sufficient promptness, white noise may be injected with excessive delay, thereby reducing the attack effectiveness and increasing the likelihood of detection through basic approaches like spectrum analysis. Moreover, prior works rely on limited-bandwidth configurations, which do not accurately reflect realistic deployment scenarios. In practical environments, resource allocations may vary dynamically across slots due to different forms of frequency hopping, further complicating timely and precise jamming operations\,\cite{963811}.
For these reasons, this paper presents a detailed study on methods for adapting the jamming process to effectively target the Physical Uplink Shared Channel (PUSCH) under realistic conditions. It addresses scenarios involving frequency hopping \cite{ristic2022frequency}, ensuring that interference remains aligned with the dynamically changing frequency allocations, as well as cases in which the bandwidth of the data transmitted over the PUSCH varies across slots. 
We note that real-time adaptation of bandwidth and central frequency is particularly challenging because of hardware-induced latency \cite{11162221} \cite{6735640}, which arises from the need to reconfigure these parameters prior to each payload transmission. In this paper, we explore techniques to mitigate such delays, allowing the proposed framework to dynamically tune its transmission settings to align with the bandwidth and central frequency of the targeted uplink resources. This adaptive approach improves the energy efficiency of the attack: by confining the interference to only the resources allocated to the target UE, STORM-RJ achieves the same disruptive effect as a wideband jammer while transmitting over a fraction of the spectrum. A non-selective jammer covering the entire system bandwidth would waste transmit power on unoccupied or irrelevant resources, whereas precise time–frequency alignment ensures that all radiated power is directed exclusively toward disrupting the intended transmission.

\section{Technical Background}
\label{sec:technical background}
\begin{figure}[!h]
    \centering
    \includegraphics[width=0.85\columnwidth]{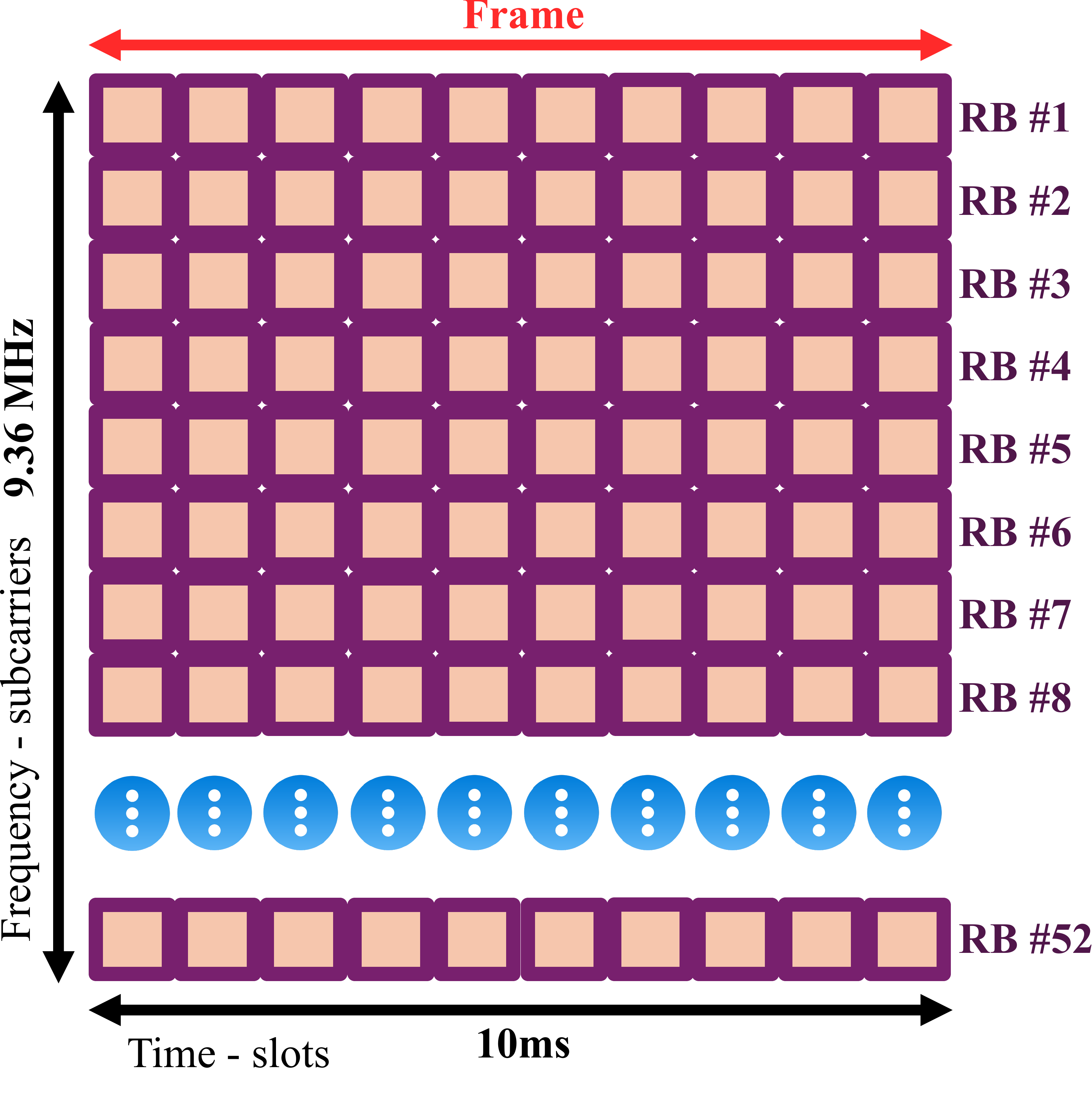}
    \caption{5G-NR time-frequency resource grid for one frame.}
    \label{fig:RA_4}
\end{figure}

A thorough understanding of how time-frequency resources are organized and scheduled in the 5G-NR air interface is essential for designing an effective jamming strategy. In 5G-NR, radio transmissions are structured in frames of 10\,ms, each composed of 10 subframes of 1\,ms. Each subframe is further divided into slots, whose duration depends on the selected \emph{numerology}, a central concept in NR denoted by $\upmu$ which defines the Subcarrier Spacing (SCS) as

\begin{equation}
\begin{alignedat}{2}
\Delta f = 15 \times 2^{\mu} \ \text{kHz},
\end{alignedat}
\label{eq:scs_and_slots_side_left}
\end{equation}
and consequently determines the Orthogonal Frequency Division Multiplexing (OFDM) symbol duration and slot length.
At the same time, the numerology also impacts the number of slots per frame as 

\begin{equation}
\text{N}^{\text{FRAME}}_{\text{slot}} = 10 \cdot 2^{\mu}
\label{eq:scs_and_slots_side}
\end{equation}
Higher numerology values correspond to larger SCS, shorter symbol duration, and reduced transmission time intervals, enabling lower latency and improving robustness to phase noise. Table~\ref{tab:numerology_table} summarizes the relationship between $\upmu$, SCS, and slot duration.
\begin{table}[ht]
\centering
\caption{Impact of Numerology on 5G-NR Frame Structure}
\label{tab:numerology_table}
\begin{tabular}{|c|c|c|c|c|}
\hline
\textbf{Numerology} $\boldsymbol{\mu}$ & \textbf{SCS (kHz)} & \textbf{Slot duration (ms)} \\
\hline
0 & 15  & 1.000 \\
1 & 30  & 0.500 \\
2 & 60  & 0.250 \\
3 & 120 & 0.125 \\
4 & 240 & 0.063 \\
5 & 480 & 0.031 \\
6 & 960 & 0.016 \\
\hline
\end{tabular}
\end{table}
When $\upmu=0$, according to\,(\ref{eq:scs_and_slots_side_left}) and to\,(\ref{eq:scs_and_slots_side}), the number of slots within a frame equals the number of subframes in the same frame\,\cite{5G-Frame_Structure}. At the opposite extreme, \text{$\upmu=6$} is employed in millimeter-wave (mmWave) deployments, where the wider subcarrier spacing provides increased robustness against Doppler spread while supporting the shorter slot durations required for low-latency transmissions \cite{flores2021flexible}\cite{TS_138.211}.
In such a system, the smallest addressable unit in the time–frequency grid is the \emph{Resource Element} (RE), which corresponds to one subcarrier in frequency over one OFDM symbol in time \cite{3gpp-ts-38.211}.
A \emph{Resource Block} (RB), instead, is defined in the frequency domain as a group of 12 consecutive subcarriers, each spaced according to the SCS and so to the selected numerology $\upmu$. Therefore, an RB spans a bandwidth of $12 \times \Delta f$. In the time domain, the RB extends over a given number of OFDM symbols, depending on the resource allocation. Fig.\,\ref{fig:RA_4} shows a 5G-NR time-frequency resource grid for one frame ($52$ RB, e.g. $9.36$\,MHz excluding guard bands).
When considering \emph{Type-A mapping} for the PUSCH, as specified in \cite{TS_138.214}, the RB allocation follows a slot-based structure. In this configuration, an RB consists of 12 contiguous subcarriers in frequency over the duration of one slot.
It is important to consider that a UE transmits a certain number of RBs within a single slot. This number may vary from slot to slot and is communicated to the UE via the UL-Grant, which is delivered by the gNB through the PDCCH. 
In a representative uplink allocation scenario, in a 5G-NR cell configured with a 10\,MHz channel bandwidth and numerology $\upmu=0$, a UE can be scheduled with up to 52\,RBs within a single slot. This value corresponds to the maximum number of resource blocks available. The corresponding occupied bandwidth is therefore
\[
52 \times 12 \times 15\,\text{kHz} = 9.36\,\text{MHz},
\]
excluding guard bands.
However, the number of RBs allocated to a UE is dynamically determined by the gNB scheduler based on multiple factors, including buffer status reports, channel quality indicators, selected modulation and coding scheme, and overall uplink load conditions \cite{mamode2022comparative}. As a result, RB assignments may vary on a slot-by-slot basis, enabling fine-grained and adaptive time--frequency resource utilization in response to traffic demand and radio channel conditions.
When the gNB sends an UL-Grant to the UE, it specifies both the number of RBs the UE is allowed to use and the exact timing for its transmission in terms of slot. 
3GPP specifications defines the time-domain resource assignment for a UE \cite{TS_138.214}. As described in \cite{9013231}, it depends on the parameter $K_2$, that represents the number of slots between the reception of the UL-Grant and the actual payload transmission on the PUSCH. 
Starting from all the information described above, it is already possible to determine both the signal bandwidth and the exact time slot in which the UE will transmit to the gNB.
Consequently, this information may be exploited by a jammer to precisely determine both the transmission timing and the bandwidth over which to transmit interference, such as white noise. This could enable a highly selective and covert jamming strategy: selective because it disrupts only the specific portion of the spectrum allocated to the target transmission and covert because it minimizes unintended spectral emissions, thereby reducing the likelihood of detection by conventional spectrum monitoring and interference analysis systems.
For example, a jammer system can obtain such sensitive information through a backdoor that can be embedded within a legitimate application or introduced via malware that conceals a covert communication channel, activated under specific conditions once installed \cite{roseline2021comprehensive}.
\begin{figure}[!h]
    \centering
    \includegraphics[width=\columnwidth]{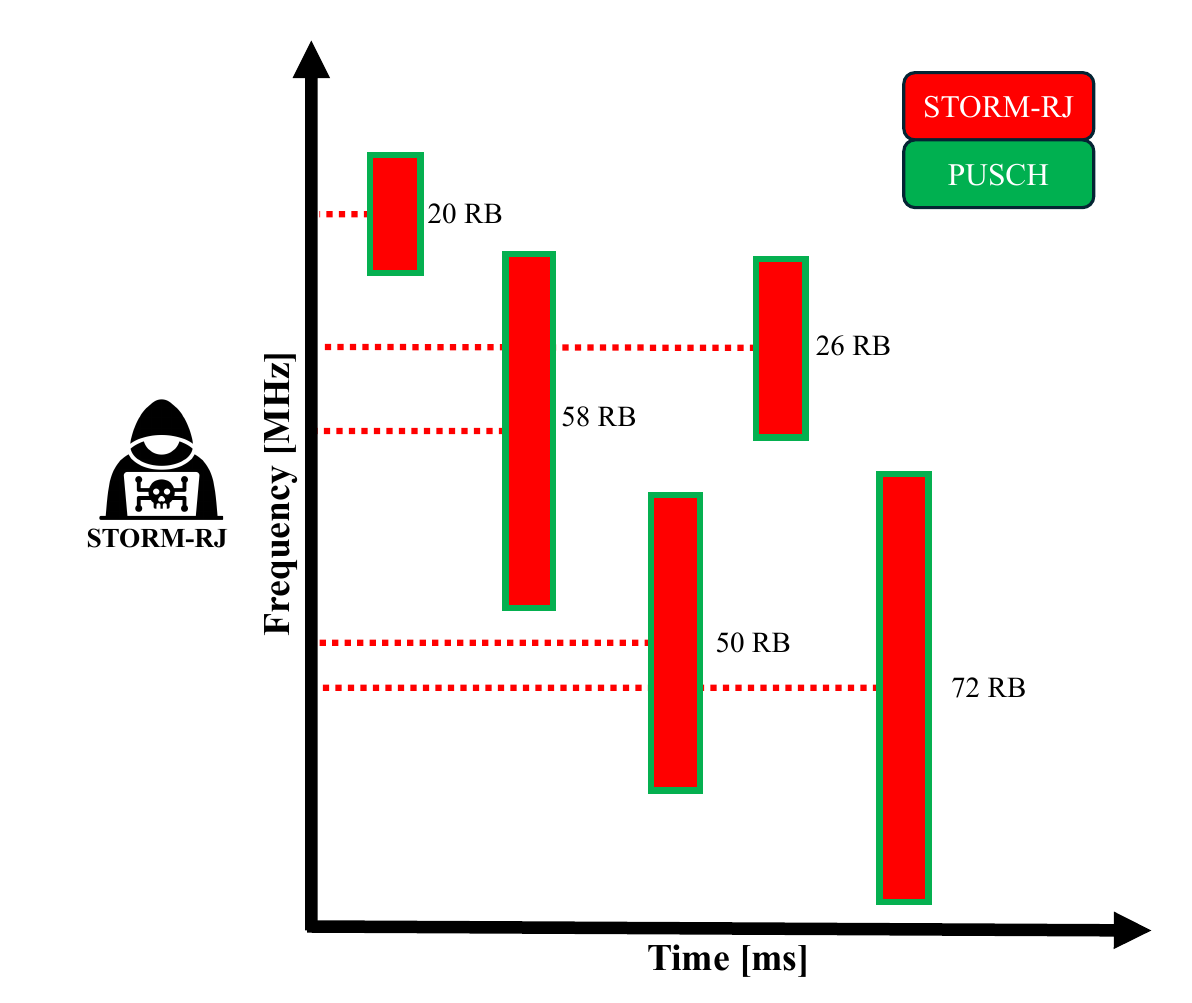}
    \vspace{-4pt}
    \caption{Time–Frequency illustration of dynamic uplink RB allocations across consecutive slots. The figure highlights how varying bandwidth assignments over time require the jammer to adapt both its center frequency and occupied bandwidth on a per-slot basis to maintain alignment with the scheduled PUSCH resources.}
    \label{fig:single_column_image}
\end{figure}
However, once the jammer acquires this scheduling information, it must compensate for four latency components in order to successfully execute the attack scenario illustrated in Fig.\,\ref{fig:single_column_image}. These components are:
\begin{enumerate}
    \item The time required by the UE to decode the DCI from the PDCCH and the subsequent transmission of the decoded UL-Grant to the jammer through the backdoor. While the former contribution is UE-dependent and not characterized in this work, the latter has been experimentally measured in \cite{STORM} at an average of approximately \SI{121}{\micro\second}.

    \item signal generation latency, namely the time required to synthesize and prepare the white-noise burst for transmission. This delay scales with the target interference bandwidth, which in turn depends on the number of RBs allocated to the UE in the UL-Grant. Wider RB allocations require broader-band noise generation, increasing processing and RF front-end reconfiguration time.

    \item latency to configure the appropriate center frequency corresponding to the scheduled PUSCH allocation. This involves translating the RB index into an absolute frequency location within the NR carrier bandwidth, taking into account the active numerology and carrier configuration. 

    \item delay required by the jammer to configure the Phase Locked Loop (PLL) to the previously computed center frequency, as this determines when the burst containing the white noise bandwidth can actually be available for transmission. 

\end{enumerate}
The cumulative effect of these latencies determines whether the interference can be precisely aligned in both time and frequency with the targeted uplink transmission.
Only after accounting for all these latency components can the feasibility of the attack be properly assessed: the jammer can successfully align its transmission with the target PUSCH only if the total accumulated delay is shorter than the time interval between the UE's DCI reception and the actual UE uplink transmission on the PUSCH. If this timing constraint is satisfied, the jammer can precisely schedule the onset of the interference in both time and frequency.
Building on this principle, we integrated the complete synchronization and reactive jamming chain leading to the development of \emph{STORM-RJ}, a multithreaded system architecture. STORM-RJ initially behaves as a standard UE, performing cell search \cite{cell_search}, time–frequency synchronization, MIB decoding, and PBCH processing to acquire the necessary system parameters from the gNB \cite{Synchronization_Procedure_in_5G_NR_Systems}. Once synchronization is established and scheduling information becomes accessible, the system switches to jamming mode.
To support this dual behavior, we implemented a custom state machine that orchestrates the full detection and synchronization pipeline — covering cell acquisition, PBCH decoding, and system information extraction — while intentionally bypassing the random access procedure. Instead of completing network attachment, the framework initializes the reactive jamming logic while maintaining slot-level synchronization with the gNB's timing reference.

\section{Methods}
\label{sec:methods}
The experimental setup consists of three separated hosts running Linux Mint 21.3, each connected to a Universal Software Radio Peripheral-SDR (USRP-SDR) from Ettus Research. 
The first host runs the UE using the srsRAN-4G framework, while the second host executes STORM-RJ, implemented by modifying the open-source free5GRAN framework\,\cite{free5gran}.
The third host executes the open-source Open5GS framework\,\cite{open5gs} to implement the Core Network (CN), responsible for handling user registration and authentication procedures within the cell. At the same time, it deploys a gNB using the srsRAN-Project framework\,\cite{srsRAN_Project}, which enables UEs to establish and maintain connectivity with the network. srsRAN-Project was preferred over srsRAN-4G as it provides native support for PUSCH frequency hopping, which is not fully implemented in srsRAN-4G due to its limited DCI format support \cite{srsRAN4G}.
Both the gNB and the UE are connected to two different Ettus Research USRP-B210 devices, each exhibiting slight frequency offsets of 0.85 and 0.83 parts per million (PPM) \cite{10597086}, respectively. For the jammer, we employed two different USRP platforms to compare their performance: a USRP-N310 and a USRP-B210.
Another USRP-N310 is included in the setup to passively monitor the PUSCH channel, using the Inspectrum tool\,\cite{inspectrum}.
To generate User Datagram Protocol (UDP) traffic between the UE and the gNB over the PUSCH, the iPerf tool\,\cite{iPerf} is used. The iPerf server runs on the gNB host, while the client is executed on the UE host.
To evaluate the practicality of targeted uplink jamming, we implemented a backdoor in our UE, specifically designed to forward the decoded DCI content directly to our jammer device. To operate the transmission of the UL-Grant from the UE to STORM-RJ, we established a client-server communication between the two devices, emulating the behavior of a backdoor capable of reliably delivering the UL-Grant to STORM-RJ.
The experiment was conducted in our laboratory, as it demonstrates that the experiment can be reliably reproduced even outside an isolated environment, bringing the setup closer to a realistic deployment scenario. In the physical arrangement of the setup, the UE is positioned 150\,cm from the gNB. The attacker is placed in between, located 50\,cm from the UE and 100\,cm from the gNB.
In our setup, communication operates in Frequency Division Duplex (FDD) mode, while the gNB was configured with an Absolute Radio Frequency Channel Number (ARFCN) of 394000, corresponding to the 3GPP n2 band \cite{n2_Band}, with center frequencies of 1970\,MHz for the downlink and 1890\,MHz for the uplink channels respectively.
A channel bandwidth ($\text{B}_{\text{CH}}$) of 20\,MHz was configured, allowing for a maximum of 106 RBs per slot. We operate with numerology $\upmu=0$, which implies an SCS of 15\,kHz. A resource mapping of type-A is adopted as well. As a consequence of these settings, the time interval between the transmission of the UL-Grant and the actual uplink payload transmission over the PUSCH channel by the UE — already defined in \ref{sec:technical background} as $\text{K}_{\text{2}}$ — is 4\,ms \cite{TS_138.214}. This interval is configured by the gNB to provide the UE with sufficient time to decode the DCI carried on the PDCCH and prepare the uplink transmission — including resource allocation, power control, and waveform generation — before the designated PUSCH slot begins \cite{TS_138.213}.
\begin{figure*}[!t]
    \centering
    \subfloat[]{%
        \includegraphics[width=0.45\textwidth]{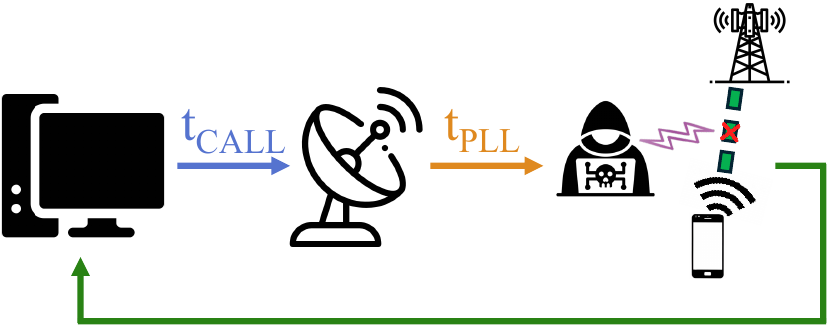}%
        \label{fig:tcall_tpll_scheme}
    }
    \hfill
    \subfloat[]{%
        \includegraphics[width=0.45\textwidth]{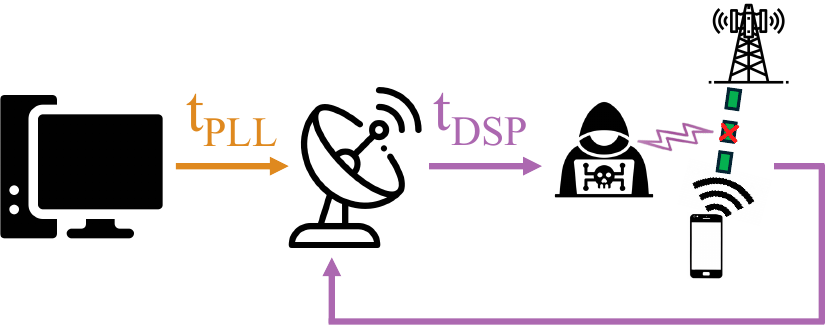}%
        \label{fig:DSP_scheme}
    }
    \caption{Schematic representations of the two center frequency tuning approaches. (a) High-level tuning: the analog LO is retuned at every UL-Grant, incurring both an API call delay $\text{t}_\text{CALL}$ and a PLL locking delay $\text{t}_\text{PLL}$. (b) Low-level tuning: the analog LO is fixed at initialization and frequency shifting is performed digitally via the NCO, introducing only the DSP processing delay $\text{t}_\text{DSP}$ on a per-grant basis.}
    \label{fig:schemes}
\end{figure*}
To ensure the effectiveness of the attack, it is essential to perform a timing analysis that quantifies the latency budget available to the jammer. 
Upon reception of each UL-Grant, STORM-RJ generates an interference burst covering the exact bandwidth allocated to the target UE. However, to further minimize the delay between successive UL-Grants, STORM-RJ was designed to pre-generate 106 interference bursts prior to the actual start of the attack, each lasting the duration of one slot and covering a bandwidth that is an integer multiple of 180\,kHz, that is the width of a single Physical Resource Block (PRB) — a group of 12 consecutive subcarriers over one slot — at 15\,kHz SCS, ranging from a minimum of 180\,kHz (single PRB) to a maximum of 19.08\,MHz (106 PRBs). At runtime, the appropriate burst is selected based on the number of PRBs indicated in the UL-Grant, enabling immediate transmission without incurring waveform generation overhead.
Secondly, we investigated two different tuning methods for configuring the USRP used in the jamming process: i) high-level tuning and ii) low-level tuning. They are subsequently described:
\begin{enumerate}
    \item \textbf{High-level tuning}. As illustrated in Fig.\,\ref{fig:tcall_tpll_scheme}, center frequency reconfiguration is performed by retuning the analog Local Oscillator (LO) of the AD9361 RF transceiver via a high-level UHD Application Programming Interface (UHD-API) call. This process requires updating the internal PLL configuration and waiting for it to lock onto the target frequency, introducing two sequential delays: the API call processing time $\text{t}_\text{CALL}$ and the PLL locking time $\text{t}_\text{PLL}$.
    
    \item \textbf{Low-level tuning}. As illustrated in Fig.\,\ref{fig:DSP_scheme}, frequency reconfiguration is performed entirely in the digital domain by adjusting the Numerically Controlled Oscillator (NCO), implemented either in the USRP-FPGA or in the internal digital mixer of the AD9361. Unlike high-level tuning, where $\text{t}_\text{CALL}$ and $\text{t}_\text{PLL}$ are incurred at every UL-Grant to retune the LO to the new target frequency, this approach requires the UHD-API call only once at initialization to fix the analog LO, and subsequently shifts the baseband spectrum digitally on a per-grant basis. This eliminates the PLL relock overhead entirely, achieving a substantially lower reconfiguration latency compared to high-level tuning.

\end{enumerate}

\subsection{Central Frequency Configuration}
\label{subsection:central-frequency-configuration}
STORM-RJ's architecture is designed to generate and radiate short bursts of white noise, whose central frequency and bandwidth dynamically depend on the radio resource allocation received from the UE's UL-Grant.
The center frequency of each transmission burst is defined according to the PRBs allocated for the uplink transmission.
Given the zero-based indices of the first and last allocated PRB ($\text{PRB}_{\text{start}}^{\text{idx}}$ and $\text{PRB}_{\text{end}}^{\text{idx}}$, both inclusive), the center frequency corresponding to the PRB group assigned to the target UE in the current slot ($f_{\text{c}}^{\text{PRB}_{x}}$) is computed as:
\begin{equation}
    f_{\text{c}}^{\text{PRB}_{x}} = f_{\text{c}}^{\text{PRB}_{\text{start}}^{\text{idx}}} + 
    \frac{\left(\text{PRB}_{\text{end}}^{\text{idx}} - \text{PRB}_{\text{start}}^{\text{idx}}\right) \cdot \text{B}_{\text{RB}}}{2}
    \label{eq:prb_center}
\end{equation}
where $\text{B}_{\text{RB}} = 180$~kHz is the bandwidth of a single RB for numerology $\upmu = 0$, 
and $f_{\text{c}}^{\text{PRB}_{\text{start}}^{\text{idx}}}$ is the center frequency of the PRB 
identified by the index $\text{PRB}_{\text{start}}^{\text{idx}}$, whose position within the 
PUSCH resource grid is determined by the UL-Grant and may correspond to any PRB in the grid.
This computation yields the center frequency corresponding to the midpoint of the allocated PRB range, enabling STORM-RJ to precisely align with the spectral location of the target resources within the 5G-NR uplink frame structure.
\begin{figure}[t]
    \centering
    \includegraphics[width=0.95\columnwidth]{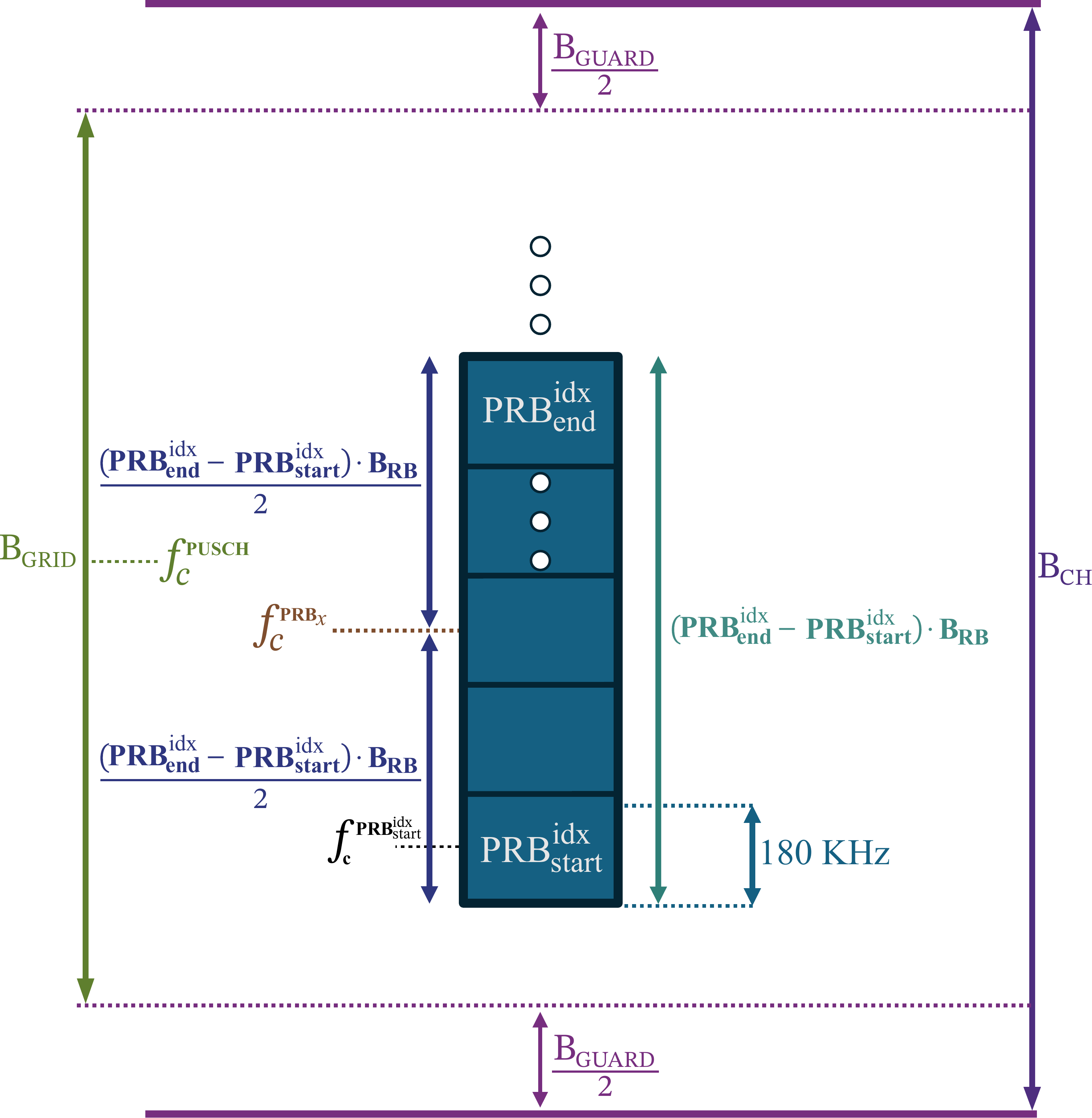}
    \caption{Graphical representation of the frequency-domain parameters used in this work, showing the spatial relationship between $\text{B}_\text{CH}$, $\text{B}_\text{grid}$, $\text{B}_\text{guard}$, and the jamming center frequency $f_{\text{c}}^{\text{PRB}_{x}}$ as a function of the allocated PRB range.}
    \label{fig:fc_prbs}
\end{figure}
In 5G-NR, the center frequency of the lowest-indexed PRB in the PUSCH resource grid is defined according to the reference grid specified in~\cite{3gpp-ts-38.104}, which outlines how the resource grid is aligned in frequency relative to the ARFCN and the corresponding carrier center frequency.
The offset between the center frequency of the PUSCH ($f_{\text{c}}^\text{PUSCH}$) and the lowest-indexed PRB in the PUSCH resource grid depends on the numerology index $\upmu$, the SCS, and the total channel bandwidth $\text{B}_\text{CH}$, as introduced in Section~\ref{sec:technical background}.
According to this specification, our setup allows for a maximum of 106 PRBs to be allocated for uplink transmission, with indices ranging from 0 to 105:
\begin{equation}
    \text{N}_\text{RB}^{\text{UL}} = 106~\text{PRB}.
    \label{eq:prb_number}
\end{equation}
Given that each PRB spans a bandwidth of 180\,kHz, the effective total bandwidth of the resource grid is:
\begin{equation}
    \text{B}_\text{grid} = \text{N}_\text{RB}^{\text{UL}} \cdot \text{B}_\text{RB} = 19.08~\text{MHz},
    \label{eq:grid_bw}
\end{equation}
where $\text{B}_{\text{grid}}$ is the usable bandwidth of the PUSCH resource grid available for uplink payload transmission. This value is slightly less than the nominal channel bandwidth of 20\,MHz, as the remaining bandwidth is reserved for guard bands at both edges of the spectrum\,\cite{3gpp-ts-38.104}. The guard band on each side is therefore:
\begin{equation}
    \frac{\text{B}_{\text{guard}}}{2} = \frac{\text{B}_\text{CH} - \text{B}_\text{grid}}{2} = 0.46~\text{MHz}.
    \label{eq:offset}
\end{equation}
The center frequency of the PRB identified by $\text{PRB}_{\text{start}}^{\text{idx}}$ can thus be expressed as:
\begin{equation}
    f_{\text{c}}^{\text{PRB}_{\text{start}}^{\text{idx}}} = f_{\text{c}}^{\text{PUSCH}} - \frac{\text{B}_\text{grid}}{2} + \left(\text{PRB}_{\text{start}}^{\text{idx}} + \frac{1}{2}\right) \cdot \text{B}_{\text{RB}}
    \label{eq:fc_prbstart}
\end{equation}
Substituting the values of our experimental setup into~(\ref{eq:fc_prbstart}) with $f_{\text{c}}^{\text{PUSCH}} = 1890$\,MHz, if we consider for instancewhere $\text{B}_{\text{RB}} = 180$~kHz is the bandwidth of a single RB for numerology $\upmu = 0$, 
and $f_{\text{c}}^{\text{PRB}_{\text{start}}^{\text{idx}}}$ is the center frequency of the PRB 
identified by the index $\text{PRB}_{\text{start}}^{\text{idx}}$, whose position within the 
PUSCH resource grid is determined by the UL-Grant and may correspond to any PRB in the grid. $\text{PRB}_{\text{start}}^{\text{idx}} = 0$, we obtain $f_{\text{c}}^{\text{PRB}_{\text{start}}^{\text{idx}}} = 1880.55$\,MHz, which corresponds to the case in which $\text{PRB}_{\text{start}}^{\text{idx}}$ coincides with the lowest-indexed PRB of the PUSCH resource grid. In general, however, $f_{\text{c}}^{\text{PRB}_{\text{start}}^{\text{idx}}}$ shifts by $\text{B}_{\text{RB}}$ for each unit increase in $\text{PRB}_{\text{start}}^{\text{idx}}$. Consequently, the last PRB in the grid, $\text{PRB}_{105}$, is centered at $f_{\text{c}}^{\text{PRB}_{105}} = 1899.45$\,MHz.
Once the center frequency corresponding to the target PRB group is determined via~(\ref{eq:prb_center}), STORM-RJ computes the appropriate frequency at which to transmit the interference burst.
To this end, STORM-RJ sets the LO to match the PUSCH carrier frequency, i.e., $f_{\text{LO}} = f_{\text{c}}^{\text{PUSCH}}$, and applies a fine digital frequency shift defined as:
\begin{equation}
    \Delta_{\text{DSP}} = f_{\text{LO}} - f_{\text{c}}^{\text{PRB}_{x}}
    \label{eq:DSP}
\end{equation}
This digital offset allows STORM-RJ to precisely align its transmitted signal with the frequency of the targeted PRB group. As a result, the actual transmission frequency is given by:
\begin{equation}
    f_{\text{TX}} = f_{\text{LO}} - \Delta_{\text{DSP}}
    \label{eq:tx_freq}
\end{equation}
The value of $f_{\text{TX}}$ is updated upon the reception of each UL-Grant. To prevent aliasing effects during digital frequency shifting, an appropriate sampling rate must be used. In our implementation, we adopted a sampling rate of $f_\text{s} = 23.04$\,MHz, which ensures safe and accurate low-level tuning within the valid frequency range.

\subsection{Bandwidth Selection and Noise Generation}
The bandwidth of each burst is directly proportional to the number of allocated PRBs, according to:
\begin{equation}
    \text{B}_{\text{burst}} = \text{N}_{\text{RB}} \cdot \text{B}_{\text{RB}}
    \label{eq:burst_bw}
\end{equation}
where $\text{N}_{\text{RB}}$ is given by the difference between $\text{PRB}_{\text{end}}$ and $\text{PRB}_{\text{start}}$.  
For each possible allocation, a pre-computed burst of complex white noise is generated through an Inverse Fast Fourier Transform (IFFT) procedure that ensures the desired spectral confinement.
The noise generation process starts by defining a frequency-domain vector $\mathbf{F}$ of length $\text{N}$, corresponding to the number of samples in a 1\,ms burst:
\begin{equation}
    \text{N} = \frac{f_s}{1000}
    \label{eq:num_samples}
\end{equation}
Where $f_s$ denotes the sampling frequency.
A set of $\text{K}$ active subcarriers is determined by the target bandwidth $\text{B}_{\text{burst}}$ and the frequency resolution $\Delta f = f_s / \text{N}$, i.e.,
\begin{equation}
    \text{K} = \left\lfloor \frac{\text{B}_{\text{burst}}}{\Delta f} \right\rfloor
    \label{eq:active_subcarriers}
\end{equation}
Random samples drawn from a zero-mean Gaussian distribution $\mathcal{N}(0,1)$ populate the frequency bins corresponding to positive spectral indices, while the remaining bins are assigned according to the Hermitian symmetry constraint to ensure a real-valued time-domain signal:
\begin{equation}
    F[-k] = F^{*}[k], \quad k = 1,2,\dots,\frac{\text{K}}{2}
    \label{eq:hermitian_symmetry}
\end{equation}

Finally, the burst waveform is computed as:
\begin{equation}
    x[n] = \frac{1}{\text{N}}\,\mathrm{IFFT}\{F[k]\}, \quad n = 0,1,\dots,\text{N}-1
    \label{eq:ifft}
\end{equation}
thus, the burst is subsequently normalized to prevent signal clipping and stored in memory for future use.
\begin{figure}[t]
    \centering
    \includegraphics[width=\columnwidth]{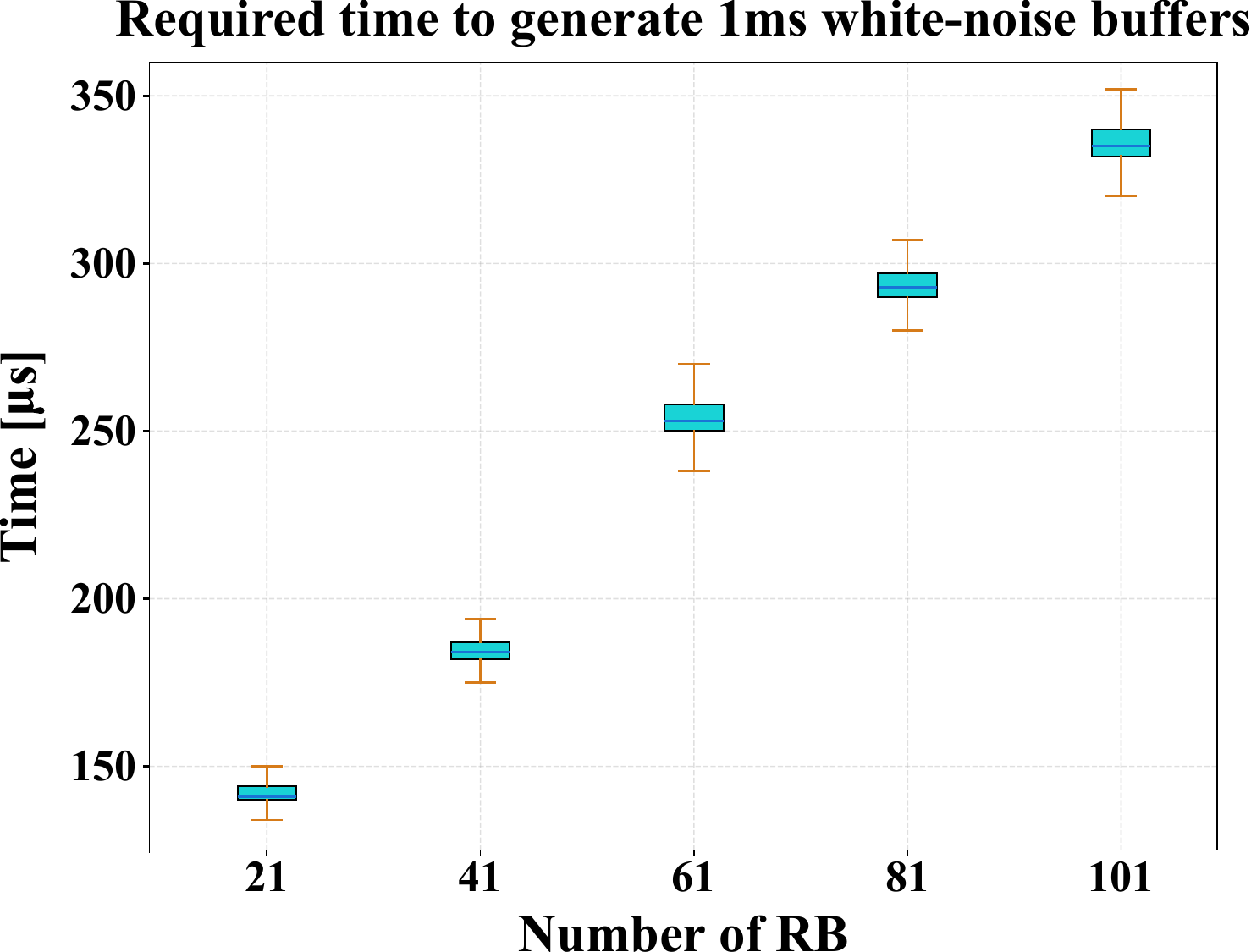}
    \vspace{-4pt}
    \caption{Average generation time required to synthesize 1\,ms white-noise bursts as a function of the number of allocated PRBs. Results are averaged over 1000 iterations. Larger bandwidth allocations increase the computational latency, motivating the adoption of a pre-buffering strategy.}
    \label{fig:burst_bandwidth}
\end{figure}
Fig.\,\ref{fig:burst_bandwidth} illustrates the generation delays associated with producing 1\,ms bursts as a function of the number of PRBs contained within each burst. An error bar plot is used to illustrate the statistics computed over 1000 iterations. 
As expected, larger burst bandwidths require longer generation time. Although the delay introduced by burst generation is relatively small, it becomes non negligible when strict timing constraints are imposed. 
To eliminate this overhead during runtime, we precompute and store all possible burst configurations in a dedicated buffer. 
This ensures that, upon receiving an UL-Grant, the corresponding pre-generated burst matching the required bandwidth can be immediately retrieved and transmitted, significantly reducing computational latency in time sensitive operations.
Upon receiving an UL-Grant and determining the UE's allocated bandwidth, STORM-RJ selects the corresponding pre generated burst and transmits it at the appropriate center frequency, as explained in \ref{subsection:central-frequency-configuration}. During execution, the selected white noise vector, corresponding to the required bandwidth, is retrieved from memory and transmitted immediately following the frequency tuning phase. The sequence of operations described above is summarized in Algorithm\,\ref{alg:STORM-RJ}, which provides a structured representation of the complete STORM-RJ workflow.

\begin{algorithm}[!hb]
\caption{STORM-RJ Workflow}
\label{alg:STORM-RJ}
\begin{algorithmic}[1]
\vspace{0.25em}
\STATE \textbf{Phase 0: Network Setup and UL Payload Generation}
\STATE Establish a standard UE--gNB connection with a compromised UE
\STATE Start uplink traffic generation using iPerf (UE $\rightarrow$ gNB)
\vspace{0.25em}
\vspace{0.25em}
\STATE \textbf{Phase 1: STORM-RJ Boot}
\STATE Start STORM-RJ, initially operating as a standard UE
\STATE Perform cell search and time--frequency synchronization with the gNB
\STATE Switch STORM-RJ to \emph{jamming mode} after synchronization
\STATE Pre-generate and store 1\,ms complex white-noise bursts for multiple bandwidth configurations
\STATE Set $\text{J}_\text{idle}$ and initialize the timer $\text{t}_{\text{last\_jam}}$
\vspace{0.25em}
\STATE \textbf{Phase 3: Reactive Jamming Loop}
\WHILE{STORM-RJ is running}

    \STATE Wait for a backdoored UL-Grant packet from the UE
    \STATE Parse $(\text{tti}_\text{tx}, \text{PRB}_\text{start}, \text{PRB}_\text{end})$ and derive the target bandwidth using (\ref{eq:burst_bw})
    
    \STATE Compute the burst center frequency associated with the allocated PRBs using (\ref{eq:fc_prbstart}) and (\ref{eq:prb_center})
    \STATE Compute the low-level tuning offset using (\ref{eq:DSP})
    \STATE Set the USRP-B210 transmit frequency according to (\ref{eq:tx_freq})
    
    \STATE Wait until the transmission instant corresponding to $\text{tti}_{\text{tx}}$, accounting for all hardware-induced and processing delays
    
    \IF{elapsed time since $\text{t}_{\text{last\_jam}} \ge \text{J}_\text{idle}$}
        \STATE Transmit the pre-buffered 1\,ms burst matching the target bandwidth
        \STATE Update $\text{t}_{\text{last\_jam}} \leftarrow$ 0
    \ELSE
        \STATE Discard the incoming UL-Grant
    \ENDIF

\ENDWHILE

\end{algorithmic}
\end{algorithm}

\begin{figure*}[!t]
    \centering
    \subfloat[]{
        \includegraphics[width=0.45\textwidth]{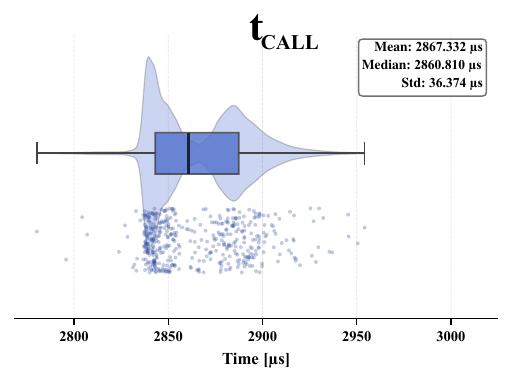}
        \label{fig:boxplot_tcall}
    }
    \hfill
    \subfloat[]{
        \includegraphics[width=0.45\textwidth]{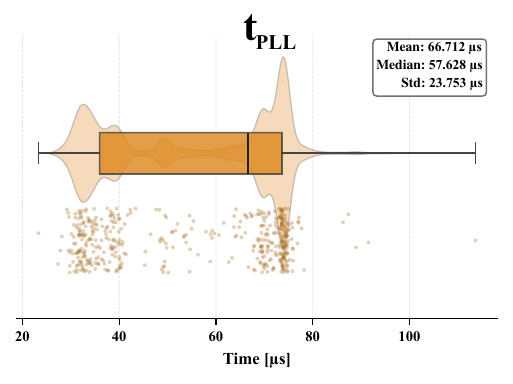}
        \label{fig:boxplot_tpll}
    }
    \caption{Measured latency distributions for high-level tuning on the USRP-B210 over 1000 repetitions. (a) Distribution of the UHD-API call delay $t_{\text{CALL}}$, showing a tightly concentrated response around \SI{2867}{\micro\second}. (b) Distribution of the PLL locking delay $t_{\text{PLL}}$, exhibiting higher variability with a mean around \SI{67}{\micro\second}. The cumulative effect of both delays approaches \SI{3}{\milli\second}, confirming the incompatibility of high-level tuning with the timing constraints of reactive jamming.}    \label{fig:tcall_tpll_distributions}
\end{figure*}

\begin{figure}[!hb]
    \centering
    \includegraphics[width=\columnwidth]{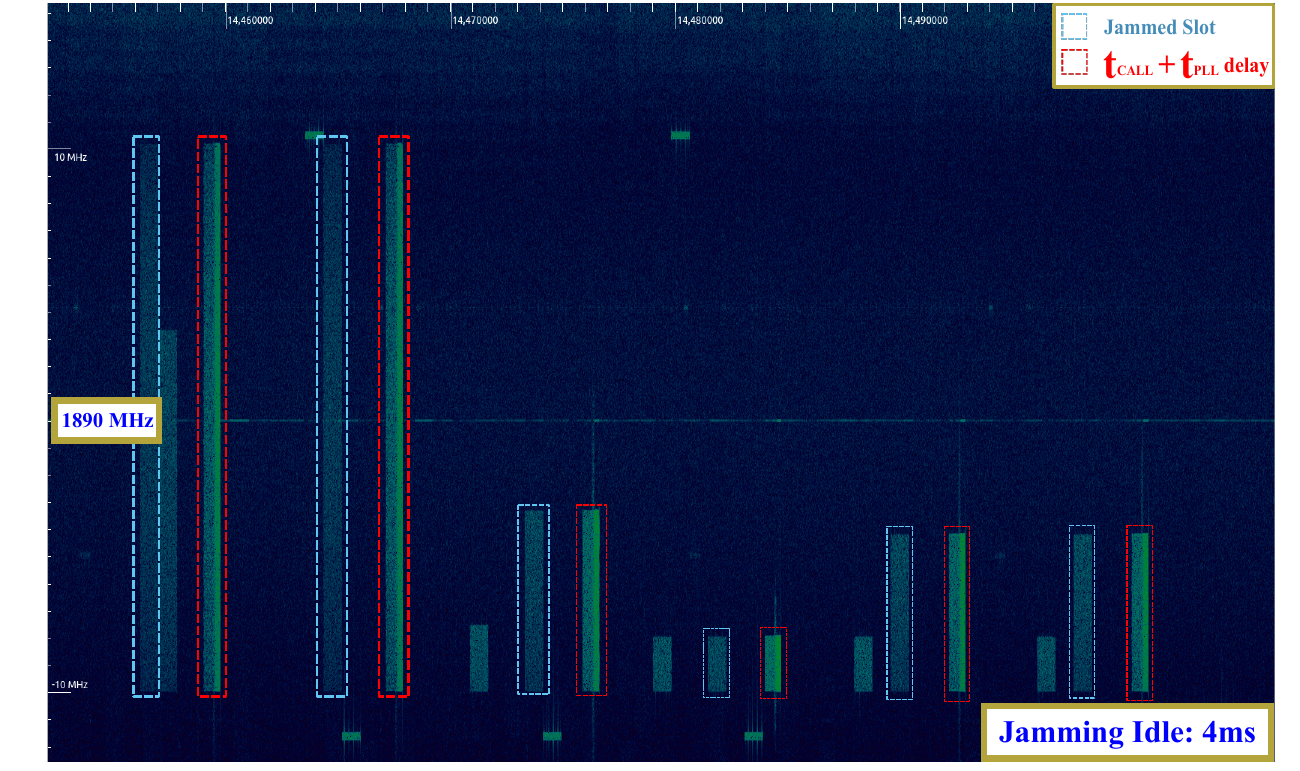}
    \vspace{-4pt}
    \caption{Spectrogram of STORM-RJ transmissions under high-level tuning ($\text{J}_\text{idle}=7\,ms$), showing the temporal offset between each UE uplink transmission and the corresponding jamming burst. The offset is a direct consequence of the
    $\text{t}_\text{CALL}+\text{t}_\text{PLL}$ retuning latency, which prevents the jammer from aligning with the intended slot. Idle slots between consecutive jamming events reflect the configured $\text{J}_\text{idle}$}
    \label{fig:inspectrum_tcall_tpll}
\end{figure}

\begin{figure*}[!t]
    \centering
    \subfloat[]{%
        \includegraphics[width=0.45\textwidth]{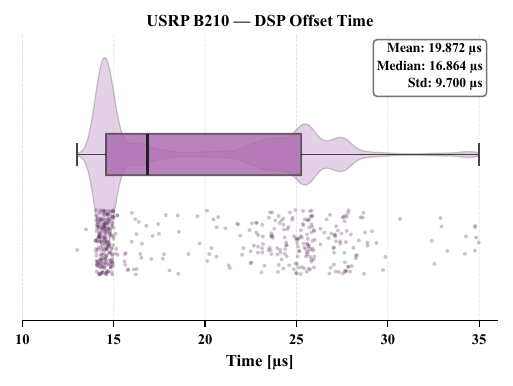}%
        \label{fig:comparison_b210}
    }
    \hfill
    \subfloat[]{%
        \includegraphics[width=0.45\textwidth]{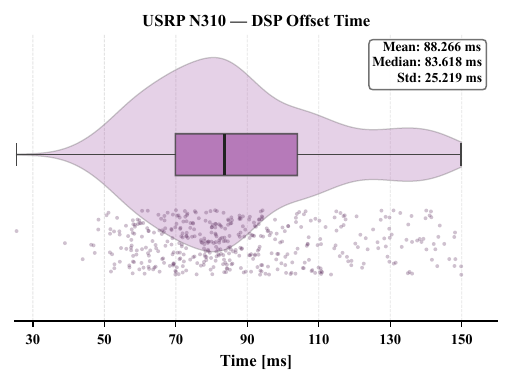}%
        \label{fig:comparison_n310}
    }
    \caption{%
    Low-level tuning evaluation and comparison across hardware platforms. (a) Distribution of DSP retuning delay for the USRP-B210 over 1000 frequency updates. (b) Equivalent measurement for the USRP-N310, highlighting substantially higher reconfiguration latency.}
    \label{fig:dsp_comparison}
\end{figure*}

\section{Results}
\label{sec:results}
In this section, we present the results obtained from the two different setups illustrated in Fig.\,\ref{fig:tcall_tpll_scheme} and Fig.\,\ref{fig:DSP_scheme}. 
Rather than the paper presented in \cite{IEEE-COINS-2025}, where the impact of selective uplink jamming on throughput degradation and radio link failure was demonstrated, this paper focuses on the engineering challenges that determine the real-world feasibility of such attacks, namely the precise time–frequency alignment of interference under dynamic resource scheduling, variable bandwidth allocation, and frequency hopping conditions. 
We define the jamming idle interval ($\text{J}_{\text{idle}}$) as the minimum time duration, expressed in milliseconds, during which STORM-RJ discards incoming UL-Grants without processing them, refraining from transmitting any jamming burst between two successive attack events. By adjusting $\text{J}_{\text{idle}}$, STORM-RJ can precisely control the temporal density of the interference, and consequently the degree of throughput degradation imposed on the target UE.
To illustrate the effect of $\text{J}_{\text{idle}}$, Fig.\,\ref{fig:inspectrum_tcall_tpll} shows an example of an attack configured with $\text{J}_{\text{idle}} = 4$\,ms. Note that $\text{J}_{\text{idle}}$ represents a lower bound on the inter-attack interval rather than a fixed periodicity: once the idle period expires, STORM-RJ resumes monitoring incoming UL-Grants and triggers the next jamming burst on the first grant received from the target UE. Since each UL-Grant refers to a future uplink slot, the actual jamming event will necessarily occur after the idle period has elapsed, with the exact timing determined by when the next UL-Grant is received and the residual reaction window available to STORM-RJ at that moment.
This behavior is governed by the USRP timer $\text{t}_\text{last\_jam}$, which is reset after each jamming event. 
A new attack is triggered only when $\text{t}_\text{last\_jam} \geq \text{J}_{\text{idle}}$ and a new UL-Grant is received, even if one or more UL-Grants arrive during the idle period. In the example shown in Fig.\,\ref{fig:inspectrum_tcall_tpll}, the payload transmission from the UE is represented by darker signals with lower power, while STORM-RJ transmission is depicted in a lighter color due to its higher power. The temporal offset visible between the UE's uplink transmission and the jamming burst is a direct consequence of the $\text{t}_\text{CALL}+\text{t}_\text{PLL}$ delays introduced by the high-level tuning procedure, which prevent STORM-RJ from completing the LO retuning process before the target slot begins. Jammer power was increased to enhance visibility and enable a direct comparison with the signals transmitted by the UE. It should be noted that in a real attack scenario, the jammer transmission power would be reduced to minimize spectral footprint and avoid detection through conventional spectrum analysis. The configurability of STORM-RJ makes it possible to select the desired level of degradation on the PUSCH channel: lower $\text{J}_\text{idle}$ values result in more frequent jamming events and, therefore, in a stronger degradation of the throughput. An example of a higher $\text{J}_\text{idle}$ configuration is shown in Fig.\,\ref{fig:final_attack}.

\subsection{High-level tuning}
The first setup we analyze is the one illustrated in Fig.\,\ref{fig:tcall_tpll_scheme}. For each received UL-Grant, STORM-RJ reconfigures the LO of the RF front-end, as previously described in Section\,\ref{sec:methods}.
Fig.\,\ref{fig:boxplot_tcall} and Fig.\,\ref{fig:boxplot_tpll} clearly show that this approach is not effective for reactive jamming, as the measured reconfiguration delays approach 3\,ms, leaving insufficient reaction time before the target PUSCH slot begins. Supporting this conclusion, Fig.\,\ref{fig:inspectrum_tcall_tpll} demonstrates how, under this setup, STORM-RJ consistently fails to jam the intended slot in time, with the jamming burst being delivered after the target slot has already elapsed.
A complete explanation of this behavior would require a detailed characterization of all latency contributions in the pipeline — including UE-side DCI processing time and potential scheduling overhead introduced by the operating system — which falls outside the scope of this work and is left as a direction for future investigation. Nevertheless, it is reasonable to conclude that the combined effect of these contributions, together with the $\approx$3\,ms incurred by the LO retuning procedure, exceeds the available reaction window, causing the jamming burst to consistently miss the intended slot.

\subsection{Low-level tuning}
In contrast to the previous setup, low-level tuning operates entirely at the DSP level and works as illustrated in Fig.\,\ref{fig:DSP_scheme}. 
Instead of reprogramming the LO of the RF front-end, a digital frequency translation is applied directly to the complex baseband signal by means of a NCO. The NCO multiplies the in-phase and quadrature (I/Q) samples by a complex exponential, as shown in \eqref{eq:dsp_shift}, effectively shifting the spectrum by a desired offset prior to digital to analog conversion.

\begin{equation}
    s'(t) = s(t) \cdot e^{j 2\pi f_{\text{DSP}} t}
    \label{eq:dsp_shift}
\end{equation}

\( f_{\text{DSP}} \) represents the digital frequency offset, while $s'(t)$ is the newly shifted signal. Since this operation is performed entirely in the digital domain, the analog LO remains fixed, and no reconfiguration or locking delay is introduced by the PLL. As a result, frequency adjustments can be performed with microsecond level latency, enabling fast and precise spectral repositioning. This makes low-level tuning particularly suitable for applications requiring real time reconfiguration or agile frequency hopping.
The valid range for \( f_{\text{DSP}} \) is inherently limited by the sampling frequency \( f_s \). According to the Shannon criterion\,\cite{1697831}, the maximum shift that can be applied without aliasing is given by \eqref{eq:dsp_nyquist}:
\begin{equation}
    |f_{\text{DSP}}| < \frac{f_s}{2}
    \label{eq:dsp_nyquist}
\end{equation}
In practice, however, this theoretical limit cannot be fully exploited due to attenuation introduced by digital filters near the spectral edges. To ensure spectral integrity and minimize distortion, the effective range is reduced as described in \eqref{eq:dsp_practical}:
\begin{equation}
    |f_{\text{DSP}}| \leq 0.45\,f_s
    \label{eq:dsp_practical}
\end{equation}
For instance, when the sampling rate is \( f_s = 23.04\,\text{MHz} \), and the LO is fixed at \( f_{\text{LO}} = 1890\,\text{MHz} \), the theoretical maximum DSP offset, is shown in \eqref{eq:dsp_example},

\begin{equation}
    |f_{\text{DSP,max}}| = \frac{f_s}{2} = 11.52\,\text{MHz}
    \label{eq:dsp_example}
\end{equation}

Nonetheless, considering the filtering constraints, the practical offset is limited to approximately \( \pm10\,\text{MHz} \). Therefore, the effective transmitted carrier frequency can be tuned digitally within the range described in \eqref{eq:dsp_tuning_range}:

\begin{equation}
    f_{\text{TX}} = f_{\text{LO}} \pm f_{\text{DSP}} \approx [1880,\,1900]\,\text{MHz}
    \label{eq:dsp_tuning_range}
\end{equation}

fixing $f_\text{LO}$ to 1890\,MHz, without requiring any future modification of the analog PLL. Operating outside this range may introduce aliasing and spectral distortion, thereby compromising signal purity.

\subsection{Evaluation results on USRP-B210 device}
\begin{figure}[!t]
    \centering
    \includegraphics[width=0.9\columnwidth]{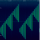}
    \vspace{-4pt}
    \caption{Spectrogram of STORM-RJ transmissions using low-level tuning on the USRP-B210, showing per-slot variation of center frequency and bandwidth across consecutive slots.}
    \label{fig:DSP_dimostrazione}
\end{figure}

\begin{figure}[!b]
    \centering
    \includegraphics[width=\columnwidth]{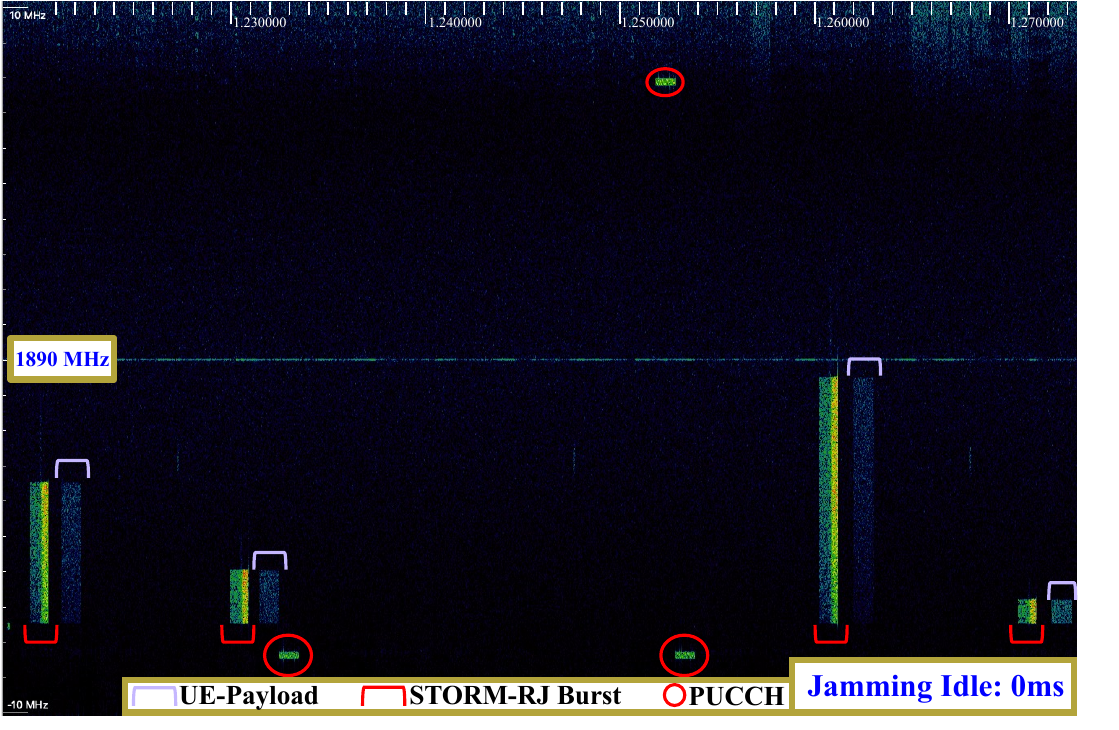}
    \vspace{-4pt}
    \caption{Spectrogram of a STORM-RJ attack ($\text{J}_\text{idle}=0$) using low-level tuning. The jamming burst is intentionally transmitted immediately upon UL-Grant decoding, without synchronization to the designated PUSCH slot, resulting in a two-slot anticipation with respect to the UE payload. Bandwidth alignment with the target PRB allocation is precise, and PUCCH transmissions remain unaffected, confirming the selectivity of the attack.}
    \label{fig:jamming_work}
\end{figure}

\begin{figure*}[ht!]
    \centering
    \includegraphics[width=\textwidth]{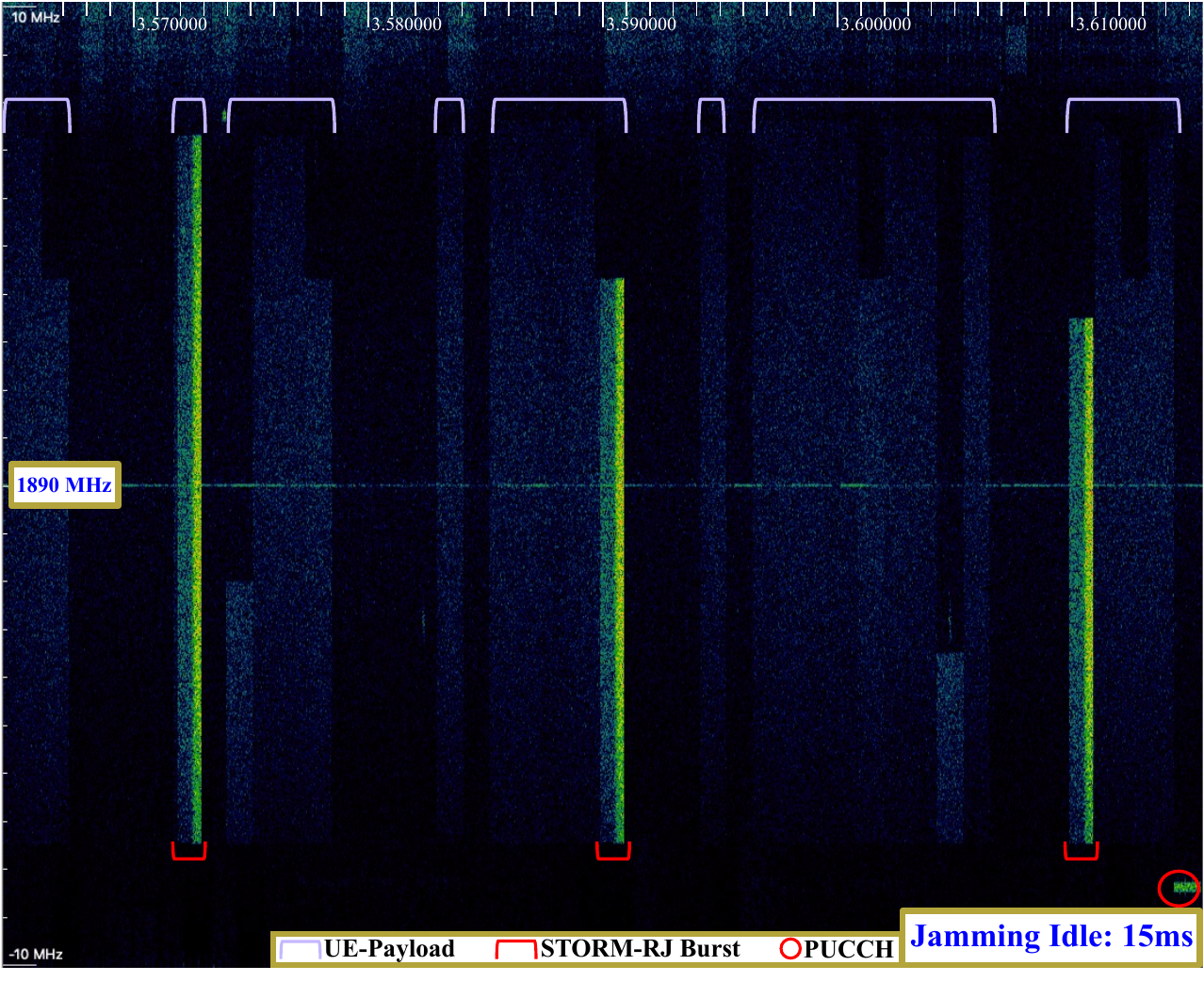}
    \vspace{-4pt}
    \caption{Spectrogram of a complete STORM-RJ attack using low-level tuning ($\text{J}_\text{idle}=15$\,ms), demonstrating precise time–frequency alignment between the jamming bursts and the target PUSCH allocations. The increased number of UE uplink transmissions is attributable to HARQ-triggered retransmissions of jammed packets. PUCCH resources remain unaffected, confirming the selectivity of the attack. $\text{J}_\text{idle}$ is set to prevent RRC release while maintaining persistent interference.}
    \label{fig:final_attack}
    \vspace{0pt}
\end{figure*}

Fig.\,\ref{fig:comparison_b210} illustrates the delay distribution obtained from 1000 center frequency changes performed using low-level tuning on the USRP-B210. The results confirm that this device achieves substantially reduced reconfiguration latency, averaging below \SI{20}{\micro\second}, making it highly suitable for reactive jamming applications.
The spectral behavior of the system is shown in Fig.\,\ref{fig:DSP_dimostrazione}, where the device successfully shifts the center frequency of the transmitted interference within the duration of a single slot, even when both bandwidth and center frequency are varied simultaneously on a per-slot basis.
Fig.\,\ref{fig:jamming_work} illustrates a preliminary attack scenario with $\text{J}_\text{idle}=0$\,ms in which STORM-RJ attempts to jam every received UL-Grant. It is worth noting that the UL-Grant contains explicit timing information specifying the exact slot in which the UE will transmit its payload.  In this figure, STORM-RJ was intentionally configured to transmit the jamming burst immediately upon UL-Grant receiving, without waiting for the designated transmission slot ($\text{tti}_\text{tx}$). As a result, the interference burst precedes the UE payload by approximately two slots. Despite this timing offset, bandwidth and central frequency alignment between the jamming burst and the target PRBs allocation is precise, confirming the effectiveness of the NCO-based frequency tuning. Notably, STORM-RJ selectively targets only the PUSCH resources, leaving the Physical Uplink Control Channel (PUCCH) transmissions unaffected.
The complete and synchronized attack scenario is presented in Fig.\,\ref{fig:final_attack}, which constitutes the core experimental result of this work. In this Figure, STORM-RJ operates with $\text{J}_\text{idle}=15$\,ms and low-level tuning, selectively jamming only the designated PUSCH slots while leaving all remaining uplink resources — including PUCCH — entirely unaffected. The precise time–frequency alignment of each jamming burst with the corresponding UE payload confirms the effectiveness of the NCO-based tuning strategy combined with the pre-buffering mechanism.
A notable consequence of the selective interference is visible in the spectrogram: the UE transmits a significantly higher number of uplink payload than expected under unperturbed conditions. 
This behavior is a direct consequence of the HARQ retransmission mechanism. When a jammed PUSCH transmission causes a Cyclic Redundancy Chech (CRC) failure at the gNB, the scheduler issues a negative Acknowledgement (NACK) and re-schedules the same Transport Block (TB) for retransmission; the receiver then performs soft combining of successive attempts to improve decoding probability. This process repeats until either the CRC succeeds or the maximum number of retransmissions is exhausted, after which the TB is permanently discarded at the MAC layer \cite{TS_138.321}.
As a result, the UE is required to serve not only its regular traffic but also the retransmissions triggered by the attack, effectively increasing the uplink load. The $\text{J}_\text{idle}$ is set sufficiently large to prevent a Radio Resource Control (RRC) Release, allowing the attack to persist without disrupting network connectivity — consistent with the stealthy design objective of STORM-RJ.
\subsection{Evaluation results on USRP-N310 device}
We also tested low-level tuning with the USRP-N310 to evaluate whether it could offer improved performance compared to the USRP-B210. 
However, as indicated in the official documentation\,\cite{usrp_n310}, this device is not optimized for fast tuning. Our experimental validation confirmed this constraint: as shown in Fig.\,\ref{fig:comparison_n310}, switching between transmissions with different center frequencies and bandwidths incurs a reconfiguration delay of approximately 88\,ms. Such latency is incompatible with the stringent timing requirements of our system, which demands rapid central frequency updates to selectively jam specific time-frequency slots with minimal delay.
Indeed, the USRP-N310 is designed to deliver high RF performance, wide instantaneous bandwidth, and robust signal quality, rather than rapid retuning. Its architecture, based on AD9371 transceivers and controlled through FPGA and UHD software, imposes tuning delays due to the internal PLL and VCO lock times required to preserve spectral purity and phase coherence. 
These delays are not the result of suboptimal implementation but stem from deliberate design trade offs aimed at maximizing signal fidelity \cite{usrp_n310_hardware}.

\subsection{Impact of $\text{J}_\text{idle}$ on UL HARQ Performance and Throughput}
\begin{figure*}[t]
    \centering
    \includegraphics[width=\textwidth]{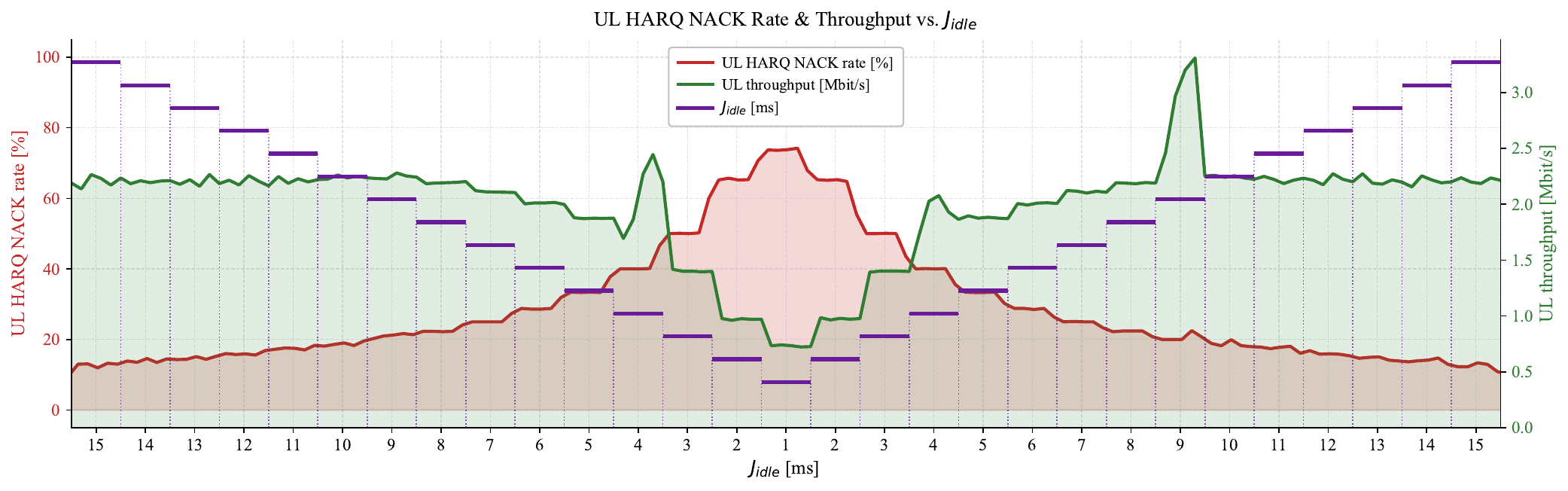}
    \vspace{-4pt}
    \caption{UL-HARQ permanent failure rate and UL-Throughput as a function of $\text{J}_\text{idle}$. Two Throughput anomalies are visible: the peak at $\text{J}_\text{idle}=4$\,ms on the descending ramp is attributed to a favorable timing alignment between the jammer period and the HARQ round-trip time, which transiently increases the average decoded TB-size per successful grant; the sharp peak at $J_{\mathrm{idle}} = 9$\,ms on the ascending ramp reflects MAC/RLC buffer drainage, as the UE flushes data accumulated during the aggressive jamming phases once channel conditions improve sufficiently.}
    \label{fig:jr_var_1_jamming_analysis}
\end{figure*}

Fig.\,\ref{fig:jr_var_1_jamming_analysis} reports the UL HARQ permanent failure rate and UL throughput as a function of $J_{\mathrm{idle}}$, measured across the full symmetric sweep from 15\,ms down to 1\,ms and back. 
For $\text{J}_\text{idle}$ values between 8 and 15\,ms, the UL throughput remains effectively constant at approximately 2.2\,Mbit/s despite a
progressively increasing permanent failure rate, which rises from 12.4\% at $\text{J}_\text{idle}=15$\,ms to 22\% at $\text{J}_\text{idle}=8$\,ms. 
This behavior is consistent with the HARQ
retransmission mechanism absorbing the interference: when a jammed transmission fails its CRC check, the scheduler re-allocates the same TB up to 3 times, and soft combining of successive attempts restores decodability without any visible throughput loss.
Below $\text{J}_\text{idle}=3$\,ms the system crosses into a qualitatively different operating point. Now the permanent failure rate reaches 50\% and throughput drops to 1.40\,Mbit/s. At $\text{J}_\text{idle}=2$\,ms, failure rate climbs to 65\% and throughput collapses to 0.97\,Mbit/s. A step before the maximum aggressiveness ($\text{J}_\text{idle}=1$\,ms) the failure rate saturates at 73.4\% and throughput reaches its minimum of 0.73\,Mbit/s, a reduction of 66.9\% relative to the $\text{J}_\text{idle}=15$\,ms baseline. 
The symmetric ramp-up path produces nearly identical values at each $\text{J}_\text{idle}$ level, confirming that the observed degradation reflects steady-state channel conditions rather than transient effects.

\section{Defensive Considerations}
\label{Defensive Considerations}
The vulnerability exploited in this study does not rely on weaknesses in the radio interface itself, but rather on the unauthorized leakage of control plane information, specifically the DCI contents, via a malicious backdoor running on a compromised UE. Although the DCI is encrypted and can only be decrypted by the intended recipient, the presence of a software level backdoor within the UE allows the attacker to extract and exfiltrate the decoded uplink scheduling information without breaking the encryption, thus enabling selective jamming with precise timing and frequency targeting.
To mitigate this risk, defensive strategies must focus on both the control and physical layers of the 5G-NR stack. We refer to these two aspects in the rest of this section. 

\subsection{Control-Plane Integrity and Anomaly Detection}
At the control-plane level, a possible mitigation strategy is the implementation of integrity verification mechanisms designed to assess the consistency between scheduled UL-Grants and the actual uplink transmission outcomes.
More specifically, the gNB can maintain statistical logs of the HARQ feedback and analyze retransmission patterns over time to detect anomalous behaviors. HARQ is a reliability mechanism used in telecommunication systems that combines forward error correction with retransmissions\cite{HARQ}. Under normal operating conditions, HARQ retransmissions are typically associated with poor radio channel conditions, which are often reflected by low Received Signal Strength Indicator (RSSI). Consequently, decoding failures tend to correlate with degraded channel measurements. In contrast, if frequent HARQ retransmissions occur despite relatively high received signal strength, this discrepancy may indicate that the decoding failure is not caused by channel attenuation but by interference, such as a reactive jamming attack targeting the scheduled resources. In parallel, real time anomaly detection mechanisms at the gNB level, potentially assisted by machine learning classifiers trained on performance indicators such as packet error rate, packet delivery ratio, or received signal strength, have been investigated as jamming detection strategies in wireless networks \cite{pirayesh2022jamming}. Such approaches may provide early warning signals of abnormal interference patterns. These systems could flag UEs whose HARQ feedback deviates from expected statistical models, or whose uplink traffic exhibits abnormal latency, packet loss, or throughput degradation under otherwise favorable radio conditions.

\subsection{Physical Layer Behavior Analysis}
Jamming attacks based on decoded DCI information produce interference that is non-random and time-synchronized with specific uplink transmissions, making them fundamentally distinguishable from ambient noise or conventional wideband jamming. This deterministic temporal and spectral correlation can be exploited by the gNB to identify interference patterns that consistently overlap with PUSCH resources allocated to a specific UE — a signature that is unlikely to arise from stochastic interference sources. Physical layer fingerprinting techniques\,\cite{11003929} can further enhance detection capabilities. Each SDR-based jammer inherently 
exhibits hardware-specific RF impairments — such as IQ imbalance, LO phase-noise, and Error Vector Magnitude (EVM) anomalies — that constitute a distinctive device fingerprint. By 
comparing the observed interference signature against the known fingerprint of the legitimate UE, the gNB can determine whether the received signal originates from an authorized transmitter. Persistent interference that fails to match the UE fingerprint, yet consistently appears during its scheduled uplink transmissions, constitutes a strong indicator of an external attacker exploiting insider scheduling information. By combining temporal correlation analysis with RF fingerprinting, network operators can substantially improve the resilience of 5G-NR systems against this class of insider-assisted attacks, where the adversary leverages partial knowledge of the scheduling state to degrade uplink performance selectively and covertly.

\section{Conclusion and Future Work}
\label{Conclusion and future work}
This paper presents an investigation into selective and stealthy jamming attacks in 5G-NR environments, with a particular focus on the PUSCH channel. We introduce STORM-RJ and demonstrate how it can efficiently target uplink transmissions by exploiting UL-Grant information obtained via a backdoor embedded in the UE. By using this information, STORM-RJ is able to align interference precisely in both time and frequency, selectively jamming the assigned RBs while remaining undetectable through conventional spectral analysis. Experimental results revealed that low-level tuning significantly outperforms high-level tuning in terms of latency, enabling center frequency adjustments with microsecond-level responsiveness and facilitating a real time attack. $\text{J}_\text{idle}$ was introduced and used to characterize the temporal distribution of the interference, illustrating how STORM-RJ can flexibly degradate PUSCH channel. 
By combining fast frequency tuning and burst preselection, the system achieves a high degree of temporal accuracy and energy efficiency.
Future research will focus on four main directions. First, we intend to extend the attack framework to target downlink communications by jamming the PDSCH. This requires ultra-low-latency execution, as the time between downlink grant (DL-Grant) reception and actual downlink transmission is less than 1\,ms. Second, we plan to evaluate the proposed attack under more realistic conditions, introducing variables such as UE mobility, multi-path fading, and concurrent transmissions from multiple UEs exploiting Arena, an open-access SDR-based testbed featuring 64 ceiling-mounted antennas and 24 symbol-level synchronized radios\cite{ARENA-Northeastern}. These scenarios will help assess the scalability, effectiveness, and robustness of STORM-RJ in practical deployments.
Third, a comprehensive end-to-end latency analysis will be conducted to characterize all timing contributions involved in the reactive jamming pipeline, including UE-side DCI processing time. 
Fourth, the security implications of the proposed attack will be investigated in the context of Open Radio Access Network (O-RAN) architectures \cite{O-RAN}. The open and standardized interfaces introduced by O-RAN may further expand the attack surface exploitable by a reactive jammer, as scheduling information could be intercepted at multiple points along the disaggregated RAN chain. Assessing the feasibility and impact of STORM-RJ in an O-RAN deployment represents a natural and timely extension of this research. 

\section*{Acknowledgment}
This work was partially supported by the European Union - Next Generation EU under the Italian National Recovery and Resilience Plan (NRRP), Mission 4, Component 2, Investment 1.3, CUPE83C22004640001, CUP E63C22002070006, CUP F83C22001690001, and CUP B53C22004050001, partnership on “Telecommunications of the Future”, PE00000001 - program “RESTART”, and Investement 7PE00000014 - CUP D33C22001300002, program SERICS.

\bibliographystyle{ieeetr}
\bibliography{bibliography}

@misc{open5gs,
  author    = {Open5GS Project},
  title     = {Open5GS},
  year      = {2024},
  howpublished = {\url{https://open5gs.org/}}
}

@misc{srsRAN4G,
  author    = {SRSRAN 4G},
  title     = {SRSRAN 4G},
  year      = {2024},
  howpublished = {\url{https://www.srsran.com/4g}}
}

@misc{srsRAN_Project,
  author    = {SRSRAN Project},
  title     = {SRSRAN 4G},
  year      = {2024},
  howpublished = {\url{https://docs.srsran.com/projects/project/en/latest/}}
}

@misc{n2_Band,
  author    = {ShareTechnote},
  title     = {5G Frequency Range and Bandwidth},
  howpublished = {\url{https://www.sharetechnote.com/html/5G/5G_FR_Bandwidth.html}}
}

@INPROCEEDINGS{10597086,
  author={Mangione, Stefano and Dino, Alessandra and Garbo, Giovanni and Croce, Daniele},
  booktitle={2024 Joint European Conference on Networks and Communications \& 6G Summit (EuCNC/6G Summit)}, 
  title={Crystal Oscillator Error Compensation in Software Defined Radios for 5G Network Testbeds}, 
  year={2024},
  volume={},
  number={},
  pages={1-5},
  doi={10.1109/EuCNC/6GSummit60053.2024.10597086}}

@misc{iPerf,
  author    = {iPerf.fr},
  title     = {iPerf3 user documentation},
  howpublished = {\url{https://iperf.fr/iperf-doc.php}}
}

@misc{TS_138.211,
  author    = {3GPP},
  title     = {TS 138.211},
  howpublished = {\url{https://www.etsi.org/deliver/etsi_ts/138200_138299/138211/18.02.00_60/ts_138211v180200p.pdf}}
}

@misc{TS_138.213,
  author    = {3GPP},
  title     = {TS 138.214},
  howpublished = {\url{https://www.etsi.org/deliver/etsi_ts/138200_138299/138213/16.02.00_60/ts_138213v160200p.pdf}}
}

@misc{TS_138.214,
  author    = {3GPP},
  title     = {TS 138.214},
  howpublished = {\url{https://www.etsi.org/deliver/etsi_ts/138200_138299/138214/16.02.00_60/ts_138214v160200p.pdf}}
}

@misc{TS_138.321,
  author    = {3GPP},
  title     = {TS 138.321},
  howpublished = {\url{https://www.etsi.org/deliver/etsi_ts/138300_138399/138321/17.14.00_60/ts_138321v171400p.pdf}}
}

@misc{free5gran,
  howpublished = {\url{https://github.com/free5G/free5GRAN}}
}

@misc{cell_search,
  howpublished = {\url{https://www.sharetechnote.com/html/5G/5G\_CellSearch.html}}
}

@misc{5G-Frame_Structure,
  howpublished ={\url{https://www.sharetechnote.com/html/5G/5G_FrameStructure.html}}
}

@article{roseline2021comprehensive,
  title={A comprehensive survey of tools and techniques mitigating computer and mobile malware attacks},
  author={Roseline, S Abijah and Geetha, S},
  journal={Computers \& Electrical Engineering},
  volume={92},
  pages={107143},
  year={2021},
  publisher={Elsevier}
}

@misc{HARQ,
  author    = {ShareTechnote},
  title     = {5G/NR-HARQ},
  howpublished ={\url{https://www.sharetechnote.com/html/5G/5G_HARQ.html}}
}

@misc{3gpp-ts-38.211,
  organization = {{3rd Generation Partnership Project (3GPP)}},
  title        = {{TS 38.211 - NR; NR and NG-RAN Overall Description}},
  howpublished = {\url{https://www.etsi.org/deliver/etsi_ts/138200_138299/138211/18.06.00_60/ts_138211v180600p.pdf?utm_source=chatgpt.com}},
  note         = {Release 17, version 17.6.0},
  year         = {2024},
}

@misc{3gpp-ts-38.104,
  organization = {{3rd Generation Partnership Project (3GPP)}},
  title        = {{TS 38.104 - NR; NR and NG-RAN Overall Description}},
  howpublished = {\url{https://www.etsi.org/deliver/etsi_ts/138100_138199/138104/16.04.00_60/ts_138104v160400p.pdf}},
  note         = {Release 17, version 17.6.0},
  year         = {2024},
}

@article{STORM,
    author = {Alaimo, Rosolino and others},
    title = {Undercover Disruption: Stealth Jamming Attacks on 5G Synchronization Stages},
    note = {In Proceedings of ITASEC-2025, (Bologna, Italy)},
    year = {2025}
}

@ARTICLE{Synchronization_Procedure_in_5G_NR_Systems,
  author={Omri, Aymen and others},
  journal={IEEE Access}, 
  title={Synchronization Procedure in 5G NR Systems}, 
  year={2019},
  volume={7},
  number={},
  pages={41286-41295},
  doi={10.1109/ACCESS.2019.2907970}}

@inproceedings{IEEE-COINS-2025,
  author={Alaimo, Rosolino and Tinnirello, Ilenia and Garlisi, Domenico},
  booktitle={2025 IEEE International Conference on Omni-layer Intelligent Systems (COINS)}, 
  title={Exploiting DCI Leakage: A Stealthy 5G Uplink Jamming Attack Using Compromised UE}, 
  year={2025},
  volume={},
  number={},
  pages={1-6},
  doi={10.1109/COINS65080.2025.11125746}
}

@misc{inspectrum,
  author       = {Michael Stapelberg},
  title        = {inspectrum: A tool for analysing captured signals},
  year         = {2015},
  howpublished = {\url{https://github.com/miek/inspectrum}},
  note         = {Accessed: 2025-12-04}
}

@misc{usrp_n310,
  author       = {{Ettus Research}},
  title        = {USRP N310 Product Page},
  year         = {2025},
  howpublished = {\url{https://www.ettus.com/all-products/usrp-n310/}},
  note         = {Accessed: 2025-12-04}
}

@misc{usrp_n310_hardware,
  author       = {{Ettus Research}},
  title        = {USRP N310 Product Page},
  year         = {2025},
  howpublished = {\url{https://files.ettus.com/manual/page_usrp_n3xx.html}},
  note         = {Accessed: 2025-12-04}
}

@article{harvanek2024survey,
  title={Survey on 5G physical layer security threats and countermeasures},
  author={Harvanek, Michal and Bolcek, Jan and Kufa, Jan and Polak, Ladislav and Simka, Marek and Marsalek, Roman},
  journal={Sensors (Basel, Switzerland)},
  volume={24},
  number={17},
  pages={5523},
  year={2024}
}

@ARTICLE{6815891,
  author={Chen, Shanzhi and Zhao, Jian},
  journal={IEEE Communications Magazine}, 
  title={The requirements, challenges, and technologies for 5G of terrestrial mobile telecommunication}, 
  year={2014},
  volume={52},
  number={5},
  pages={36-43},
  doi={10.1109/MCOM.2014.6815891}}

@ARTICLE{963811,
  author={Kostic, Z. and Maric, I. and Wang, X.},
  journal={IEEE Journal on Selected Areas in Communications}, 
  title={Fundamentals of dynamic frequency hopping in cellular systems}, 
  year={2001},
  volume={19},
  number={11},
  pages={2254-2266},
  doi={10.1109/49.963811}}

@ARTICLE{1697831,
  author={Shannon, C.E.},
  journal={Proceedings of the IRE}, 
  title={Communication in the Presence of Noise}, 
  year={1949},
  volume={37},
  number={1},
  pages={10-21},
  doi={10.1109/JRPROC.1949.232969}}

@INPROCEEDINGS{9013231,
  author={Patriciello, Natale and Lagen, Sandra and Giupponi, Lorenza and Bojovic, Biljana},
  booktitle={2019 IEEE Global Communications Conference (GLOBECOM)}, 
  title={The Impact of NR Scheduling Timings on End-to-End Delay for Uplink Traffic}, 
  year={2019},
  volume={},
  number={},
  pages={1-6},
  doi={10.1109/GLOBECOM38437.2019.9013231}}

@article{ristic2022frequency,
  title={Frequency hopping spread spectrum: History, principles and applications},
  author={Risti{\'c}, Vladimir B and Todorovi{\'c}, Branislav M and Stojanovi{\'c}, Nenad M},
  journal={Vojnotehnicki glasnik/Military Technical Courier},
  volume={70},
  number={4},
  pages={856--876},
  year={2022},
  publisher={University of Defence}
}

@INPROCEEDINGS{11162221,
  author={Laskos, Christos and Zubow, Anatolij and Dressler, Falko},
  booktitle={2025 IEEE International Conference on Communications Workshops (ICC Workshops)}, 
  title={Latency Analysis of SDR-based Experimental C-RAN / O-RAN Systems}, 
  year={2025},
  volume={},
  number={},
  pages={893-898},
  doi={10.1109/ICCWorkshops67674.2025.11162221}}

@INPROCEEDINGS{6735640,
  author={Truong, Nguyen B. and Suh, Young-Joo and Yu, Chansu},
  booktitle={MILCOM 2013 - 2013 IEEE Military Communications Conference}, 
  title={Latency Analysis in GNU Radio/USRP-Based Software Radio Platforms}, 
  year={2013},
  volume={},
  number={},
  pages={305-310},
  doi={10.1109/MILCOM.2013.60}}

@ARTICLE{9733393,
  author={Pirayesh, Hossein and Zeng, Huacheng},
  journal={IEEE Communications Surveys \& Tutorials}, 
  title={Jamming Attacks and Anti-Jamming Strategies in Wireless Networks: A Comprehensive Survey}, 
  year={2022},
  volume={24},
  number={2},
  pages={767-809},
  doi={10.1109/COMST.2022.3159185}}

@INPROCEEDINGS{9031175,
  author={Arjoune, Youness and Faruque, Saleh},
  booktitle={2020 10th Annual Computing and Communication Workshop and Conference (CCWC)}, 
  title={Smart Jamming Attacks in 5G New Radio: A Review}, 
  year={2020},
  volume={},
  number={},
  pages={1010-1015},
  doi={10.1109/CCWC47524.2020.9031175}}

@ARTICLE{10186886,
  author={Flores, Maya E. and Poisson, Devon D. and Stevens, Colin J. and Nieves, Adriyel V. and Wyglinski, Alexander M.},
  journal={IEEE Access}, 
  title={Implementation and Evaluation of a Smart Uplink Jamming Attack in a Public 5G Network}, 
  year={2023},
  volume={11},
  number={},
  pages={75993-76007},
  doi={10.1109/ACCESS.2023.3296701}}

@article{pirayesh2022jamming,
  title={Jamming attacks and anti-jamming strategies in wireless networks: A comprehensive survey},
  author={Pirayesh, Hossein and Zeng, Huacheng},
  journal={IEEE communications surveys \& tutorials},
  volume={24},
  number={2},
  pages={767--809},
  year={2022},
  publisher={IEEE}
}

@article{mamode2022comparative,
  title={Comparative analysis of scheduling algorithms in 5G uplink transmission},
  author={Mamode, M and Fowdur, Tulsi Pawan},
  journal={Journal of Engineering Research and Sciences},
  year={2022}
}

@ARTICLE{11003929,
  author={Zhang, Junqing and Ardizzon, Francesco and Piana, Mattia and Shen, Guanxiong and Tomasin, Stefano},
  journal={IEEE Transactions on Information Forensics and Security}, 
  title={Physical Layer-Based Device Fingerprinting for Wireless Security: From Theory to Practice}, 
  year={2025},
  volume={20},
  number={},
  pages={5296-5325},
  doi={10.1109/TIFS.2025.3570118}}

@article{chen2020design,
  title={Design and implementation of initial cell search in 5G NR systems},
  author={Chen, Fatang and Li, Xiu and Zhang, Yun and Jiang, Yanan},
  journal={China Communications},
  volume={17},
  number={5},
  pages={38--49},
  year={2020},
  publisher={IEEE}
}

@article{flores2021flexible,
  title={Flexible numerology in 5G NR: Interference quantification and proper selection depending on the scenario},
  author={Flores de Valgas, Josue and Monserrat, Jose F and Arslan, H{\"u}seyin},
  journal={Mobile Information Systems},
  volume={2021},
  number={1},
  pages={6651326},
  year={2021},
  publisher={Wiley Online Library}
}

@ARTICLE{O-RAN,
  author={Polese, Michele and Bonati, Leonardo and D’Oro, Salvatore and Basagni, Stefano and Melodia, Tommaso},
  journal={IEEE Communications Surveys \& Tutorials}, 
  title={Understanding O-RAN: Architecture, Interfaces, Algorithms, Security, and Research Challenges}, 
  year={2023},
  volume={25},
  number={2},
  pages={1376-1411},
  doi={10.1109/COMST.2023.3239220}}

@article{ARENA-Northeastern,
  title={Arena: A 64-antenna SDR-based ceiling grid testing platform for sub-6 GHz 5G-and-Beyond radio spectrum research},
  author={Bertizzolo, Lorenzo and Bonati, Leonardo and Demirors, Emrecan and Al-Shawabka, Amani and D’Oro, Salvatore and Restuccia, Francesco and Melodia, Tommaso},
  journal={Computer Networks},
  volume={181},
  pages={107436},
  year={2020},
  publisher={Elsevier}
}
\end{document}